\documentclass[fleqn,usenatbib]{mnras}

\usepackage{newtxtext,newtxmath}

\usepackage[T1]{fontenc}

\DeclareRobustCommand{\VAN}[3]{#2}
\let\VANthebibliography\thebibliography
\def\thebibliography{\DeclareRobustCommand{\VAN}[3]{##3}\VANthebibliography}

\usepackage{graphicx}	
\usepackage{amsmath}	
\usepackage{CJKutf8}
\newcommand{\chinese}[1]{\begin{CJK}{UTF8}{gbsn}\normalfont #1\end{CJK}} 

\title[Star formation concentration and quenching]{Star formation concentration as a tracer of environmental quenching in action: a study of the {\sc Eagle} and {\sc C-Eagle} simulations}

\author[D. Wang et al.]{Di Wang (\chinese{王迪}),$^{1,2}$\thanks{E-mail: di.wang@sydney.edu.au}
Claudia D.~P. Lagos,$^{2,3,4}$
Scott M. Croom,$^{1,2}$
Ruby J. Wright,$^{2,3}$
Yannick M. Bahé,$^{5,6}$ 
\newauthor
Julia J. Bryant,$^{1,2}$
Jesse van de Sande,$^{1,2}$
and Sam P. Vaughan,$^{2,7,8,9}$
\\
$^{1}$Sydney Institute for Astronomy (SIfA), School of Physics, The University of Sydney, NSW, 2006, Australia\\
$^{2}$ARC Centre of Excellence for All Sky Astrophysics in 3 Dimensions (ASTRO 3D)\\
$^{3}$International Centre for Radio Astronomy Research (ICRAR), M468, University of Western Australia, 35 Stirling Hwy, Crawley, WA 6009, Australia\\
$^{4}$Cosmic Dawn Center (DAWN), Denmark.\\
$^{5}$Leiden Observatory, Leiden University, PO Box 9513, NL-2300 RA Leiden, The Netherlands \\
$^{6}$Institute for Computational Cosmology, Durham University, South Road, Durham DH1 3LE, UK\\
$^{7}$School of Mathematical and Physical Sciences, Macquarie University, NSW 2109, Australia\\
$^{8}$Astronomy, Astrophysics and Astrophotonics Research Centre, Macquarie University, Sydney, NSW 2109, Australia\\
$^{9}$Centre for Astrophysics and Supercomputing, School of Science, Swinburne University of Technology, Hawthorn, VIC 3122, Australia\\
}

\date{Accepted XXX. Received YYY; in original form ZZZ}

\pubyear{2023}

\begin{document}
\label{firstpage}
\pagerange{\pageref{firstpage}--\pageref{lastpage}}
\maketitle

\begin{abstract} We study environmental quenching in the {\sc Eagle}/{\sc C-Eagle} cosmological hydrodynamic simulations over the last 11 Gyr (i.e. $z=0-2$). The simulations are compared with observations from the SAMI Galaxy Survey at $z=0$. We focus on satellite galaxies in galaxy groups and clusters ($10^{12}\,\rm M_{\odot}$ $\lesssim$ $M_{200}$ < $3 \times 10^{15}\, \rm M_{\odot}$). A star-formation concentration index [$C$-index $= \log_{10}(r_\mathrm{50,SFR} / r_\mathrm{50,rband})$] is defined, which measures how concentrated star formation is relative to the stellar distribution. {Both {\sc Eagle}/C-{\sc Eagle} and SAMI show a higher fraction of galaxies with low $C$-index in denser environments at $z=0-0.5$. Low $C$-index galaxies are found below the SFR-$M_{\star}$ main sequence (MS), and display a declining specific star formation rate (sSFR) with increasing radii, consistent with ``outside-in'' environmental quenching. Additionally, we show that $C$-index can be used as a proxy for how long galaxies have been satellites. These trends become weaker at increasing redshift and are absent by $z=1-2$. We define a quenching timescale $t_{\rm quench}$ as how long it takes satellites to transition from the MS to the quenched population.} We find that simulated galaxies experiencing ``outside-in'' environmental quenching at low redshift ($z=0\sim0.5$) have a long quenching timescale (median $t_{\rm quench}$ > 2 Gyr). The simulated galaxies at higher redshift ($z=0.7\sim2$) experience faster quenching (median $t_{\rm quench}$ < 2Gyr). At $z\gtrsim 1-2$ galaxies undergoing environmental quenching have decreased sSFR across the entire galaxy with no ``outside-in'' quenching signatures and a narrow range of $C$-index, showing that on average environmental quenching acts differently than at $z\lesssim 1$.

\end{abstract}




\section{Introduction}

The way star formation proceeds and ceases in galaxies is a major area of research in astrophysics. 
{By studying the mechanisms that quench star formation in galaxies, we can gain insight into the processes that shape the Universe around us. Additionally, linking observational information with simulation results allows us to better understand the complex processes involved in galaxy quenching and to make more accurate predictions about the future evolution of galaxies.}

Based on their star formation rate (SFR), we usually distinguish between galaxies with ongoing star formation, generally referred to star-forming (SF) galaxies, and those that are no longer forming stars, which are referred to passive galaxies. SF galaxies and passive galaxies show a clear bimodality in the SFR-stellar mass diagram \citep[e.g.][]{Abazajian_2009, Renzini_2015, Katsianis2020}. 
Many physical processes can lead to galaxy quenching (cessation of star formation), but they are broadly grouped into those that are internal to galaxies (and that correlate with the galaxy stellar mass), and those external to galaxies, which are usually related to the environments in which galaxies reside \citep[e.g.][]{Peng_2010,Peng_2012}.

When galaxies reside in dense environments such as groups and clusters, there are mechanisms that can contribute to a reduction in star formation rate and induce quenching. Such mechanisms include: ram pressure-stripping (RPS), whereby satellite gas can be stripped by the surrounding dense intracluster medium (ICM) of the host halo \citep[e.g.][]{Gunn1972, Abadi1999}; strangulation, where a satellite is cut off from gas accretion \citep[e.g.][]{Peng2015}; as well as viscous stripping or tidal stripping \citep[e.g.][]{Larson_1980, Balogh2000}. While environmental processes can act to quench galaxies, the opposite can also take place, with star formation being enhanced in some cases where there are pair-type encounters between galaxies \citep[e.g.][]{Hernquist1989,For2021}.

Early research using single-fiber surveys has revealed many correlations between the environment of a galaxy and its star-formation activity. The fraction of passive galaxies has been found to increase with environmental density  \citep[e.g.][]{Goto2003, Wijesinghe_2012}. \citet{Davies2019}, using the Galaxy And Mass Assembly survey \citep[GAMA, e.g.][]{Driver2009,Driver2011,Liske2015}, show that even at fixed stellar mass, the fraction of quenched satellite galaxies grows as the halo mass increases. Using the Sloan Digital Sky Survey \citep[SDSS, e.g.][]{York_2000}, many other authors (e.g. \citealt{Balogh2000,Ellingson2001,Peng_2012, Wetzel_2013,Renzini_2015}) have pointed to environmental processes being the primary responsible for quenching of galaxies in groups and clusters.

With the development of Integral Field Spectroscopy (IFS), we can now access spatially resolved spectra for thousands of galaxies; meaning that star formation, stellar ages, stellar metallicities, and kinematics can be measured for different parts of a galaxy. 
IFS surveys such as the Sydney-Australian Astronomical Observatory Multi-object Integral Field Spectrograph Galaxy Survey \citep[SAMI, e.g.][]{Croom2012, Croom_2021} and the SDSS-IV Mapping Nearby Galaxies at Apache Point Observatory survey \citep[MaNGA, e.g.][]{Bundy2014} have recently enabled resolved investigations of star-formation quenching in galaxies \citep[e.g.][]{Goddard2017, Ellison2018, Spindler2018, Owers2019}. Using galaxies from the SAMI galaxy survey, \citet[]{Schaefer2017} and \citet{Medling2018} found that galaxies in denser environments show decreased sSFR in their outer regions compared to isolated galaxies, consistent with environmental quenching. \citet[]{Oh_2018}, using SAMI data, found galaxies may experience environmental effects before galaxies fall into a dense enough environment like the cluster centre.

The concentration of star formation relative to the stellar continuum has been used as a smoking gun for environmental quenching. Since environmental mechanisms typically work in an outside-in manner (e.g. stripping of the tenuous, poorly-bound gas on the outskirts of satellites) it is thought that in the quenching phase, any remaining star formation in a satellite galaxy is likely to be constricted to the inner-most regions. 
At $z \sim 0$, \citet{schaefer2019} studied the concentration of star formation with the ratio of half-light radius of H$\alpha$ ($r_\mathrm{50,H\alpha}$) and half-light radius of the $r$-band continuum ($r_\mathrm{50,cont}$) and defined concentration of star formation ($C$-index) as $\log_{10}(r_\mathrm{50,H\alpha}/r_\mathrm{50,cont})$. They used SAMI group galaxies in the GAMA regions and found that galaxies in groups of masses $10^{12.5-14}M_{\odot}$ have more centrally concentrated galaxies ($C$-index < $-0.2$, 29$\pm$7 per cent) compared to field galaxies (4$\pm$4 per cent), indicative of ``outside-in'' quenching. \citet{Wang2022} extended this work 
used 
the full SAMI data and found the fraction of galaxies with concentrated star formation increases with halo mass. Galaxies with concentrated star formation in groups ($M_{200}$ in $10^{12.5-14}M_{\odot}$) have older discs compared to galaxies with extended star formation. Those galaxies may undergo partial RPS, which is galaxies are partially stripped of gas by ram-pressure in high-mass groups, suppressing star formation in the outer disc, while central star formation remains unaffected. They also found galaxies with concentrated star formation in clusters ($M_{200} > 10^{14}M_{\odot}$) have similar radial age profiles to ungrouped galaxies (not classified in a group galaxy in the GAMA Galaxy Group Catalogue version 10), suggesting quenching process must be rapid. At higher redshifts, \citet[]{Vaughan2020} show that cluster galaxies at z$\sim$0.5 have $26\pm 12$\ per cent smaller H$\alpha$-to-stellar continuum size ratios than coeval field galaxies, which is broadly consistent with \citet{Wang2022}, who found galaxies in clusters have lower $C$-index comparing to field galaxies. At $z \sim$ 1, \citet[]{Matharu2021} used the $r_\mathrm{50,H\alpha}/r_\mathrm{50,cont}$ ratio for SF cluster galaxies with the Wide Field Camera 3 (WFC3) G141 grism on the \textit{Hubble Space Telescope} (\textit{HST}), they found the ratio has no significant difference between cluster galaxies and field galaxies but a higher fraction of quenched galaxies in clusters suggesting environmental quenching. 

In parallel to the progress in observations above, theoretical models of galaxy formation have made significant contributions to our understanding of quenching in galaxies. Early research based on semi-analytic models of galaxy formation (SAMs) predicted strangulation to be a dominant quenching process in galaxy groups and clusters \citep[e.g.][]{Somerville2008a,Lagos2018}. Later, some SAMs added physical models for halo and ISM gas stripping and found simulation results to be more consistent with observations \citep[e.g.][]{Tecce2010, Lagos2014, Stevens2017, Cora2018, Xie2020}. Studies using cosmological hydrodynamical simulations predict that RPS is likely to dominate the quenching of galaxies in clusters and play a less important role in galaxy groups \citep[][]{Bahe2014, Marasco2016, Lotz2019}. Numerical simulations have also shown that other environmental effects can play an important role, such as group pre-processing \citep[e.g.][]{Fujita2004, Bahe2013, Ayromlou2019,Donnari2020, Ayromlou2021}, satellite-satellite interactions \citep{Marasco2016} and partial RPS. For the latter, \citet[]{Steinhauser2016} found that the complete gas stripping of discs only happens in extreme cases, and partial RPS can even lead to a star-formation enhancement in the galaxy centre.

\citet[]{Oman2021} compared the SDSS satellite galaxies in groups and clusters with orbital information from $N$-body simulations with an analytic model and found that galaxies are best described by a ‘delayed-then-rapid’ quenching scenario, where galaxies are consistent with continual star formation for a few Gyrs after becoming satellites, to then shut-off their star formation in only a hundred Myr (see also \citealt{Wetzel_2013}). More recently, \citet[]{Wright2022} used the Evolution and Assembly of GaLaxies and their Environments ({\sc Eagle}) simulation and found that both RPS and strangulation play an important role in quenching satellite galaxies across a range of stellar and halo masses, with quenching timescales varying widely depending on the satellite's orbit, mass and characteristics of the host halo. According to \citet[]{Wright2022}, the rapid quenching phase takes place right after the first pericentric passage for low-mass galaxies, while massive galaxies require two or even three pericentric passages to enter the rapid quenching phase. The question is how to connect these theoretical results with observational properties of galaxies. In other words, which observations indicate varying quenching timescales, and how different do they appear at different cosmic times?

In this work, we explore how the $C$-index traces the quenching of galaxies in different environments and throughout cosmic time, using the {\sc Eagle}/C-{\sc Eagle} simulations. Our first goal is to investigate the usefulness of $C$-index as a star-formation distribution indicator probing how star-formation quenching happens in simulations and observations at $z$=0, by directly contrasting the simulations with observational results from the SAMI IFS data. Our second goal is to use the {\sc Eagle} simulation to find how the $C$-index relates to the evolving quenching timescales of satellite galaxies from $z=0$ to $z=2$, which covers the main epoch of environmental quenching.

The paper is arranged as follows: we present a brief description of the {\sc Eagle}/C-{\sc Eagle}/SAMI data used in this work, and how we measure the $C$-index, satellite galaxy orbits and quenching timescales in $\S$~\ref{Esec:methord}. In $\S$~\ref{Esec:z0}, we discuss the main outcomes of our research at $z=0$. In $\S$~\ref{Esec:quenchinghighz}, we discuss how the $C$-index and its relation to environmental quenching evolves from $z=0$ to $z=2$. In $\S$~\ref{Esec:discussion}, we discuss our findings and $\S$~\ref{Esec:conclusion} presents a short summary and conclusions.

\section{Methods}\label{Esec:methord}

\subsection{Cosmological Simulations}

\begin{table}
	\centering
	\caption{The number of SAMI, {\sc Eagle}/C-{\sc Eagle} satellite galaxies in each halo mass bin. The halo mass bins are Low-Mass Groups (LMGs, $M_{200}$ $\leq 10^{12.5} M_{\odot}$), Intermediate-Mass Groups (IMGs, $10^{12.5} M_{\odot}< M_{200} \leq 10^{13.5} M_{\odot}$), High-Mass Groups (HMGs, $M_{200}$ > $10^{13.5} M_{\odot}$), and Clusters ($M_{200}$ > $10^{14} M_{\odot}$). The SAMI and {\sc Eagle} cluster galaxies are selected from HMGs with $M_{200}$ > $10^{14} M_{\odot}$ only. }
	\label{tab:number_halo}
	\begin{tabular}{lcccc} 
		\hline
		$ $ & LMG & IMG & HMG & Cluster \\
		$ $  & ${\le} 10^{12.5}M_{\odot}$ & $ 10^{12.5-13.5}M_{\odot}$ &${>} 10^{13.5}M_{\odot}$&${>} 10^{14}M_{\odot}$\\
		\hline
		SAMI & 138 & 77 & 154 & 129 \\
		{\sc Eagle} & 906 & 1175 & 523 & 209 \\
		C-{\sc Eagle} & $ - $ &  $- $ &  $- $ &  989 \\
		\hline
	\end{tabular}
\end{table}

Below we introduce the {\sc Eagle} and Cluster-{\sc Eagle} cosmological simulations as well as parameters obtained from the simulation.

\begin{table*}
	\centering
	\caption{We list the {\sc Eagle} parameters used in this work. $R_{\rm closest-approach}$, $R_{\rm current}$, $N$-pericentre, $t_\mathrm{dep}$, and $t_\mathrm{sat}$ are parameters from \citet[]{Wright2022}. }
	\label{tab:eagle_parameter}
	\begin{tabular}{lcc} 
		\hline
		Parameter  &  Unit &  Description \\
		\hline
		$R_{\rm closest-approach}$ & kpc & The closest distance between the satellite galaxy and the host halo centre, the galaxy has had \\
		$R_{\rm current}$ & kpc & The current distance between the satellite and the host halo centre  \\
		$N$-pericentre & - & The number of pericentric passages a satellite galaxy has had \\
		\hline
		$t_\mathrm{sat}$ & Gyr & Lookback-time at which the satellite galaxy became part of its current host halo  \\
		$t_\mathrm{dep}$ & Gyr & Gas depletion time, assuming that the current ISM inflow, outflow rates, and SFRs remain constant  \\
		$t_\mathrm{last-central}$ & Gyr & Lookback-time at which the satellite galaxy was last a central galaxy \\
		$t_\mathrm{last-SF}$ & Gyr & Lookback-time at which the galaxy dropped below a log$_{10}$(sSFR/yr$^{-1}$) $=$ $-11.25$ \\
		$\delta t_\mathrm{last-central}$ & Gyr & $t_\mathrm{last-central}$ - t$_{z}$, how long the galaxy has been a satellite \\
		$t_\mathrm{quench}$ & Gyr & how long galaxies take to transition from the main sequence to the quenched population in the SFR-M$_{\star}$ plane \\
		\hline
            $\Delta$MS &  dex  & {The distance between sSFR and sSFR-redshift MS in log space} \\
            \hline
	\end{tabular}
\end{table*}

\subsubsection{The {\sc Eagle} and Cluster-{\sc Eagle} simulation} \label{Esec:sample_selection}

The {\sc Eagle} (Evolution and Assembly of GaLaxies and their Environments) simulation suite \citep[][]{Schaye2015,Crain_2015} consists of a collection of cosmological hydrodynamical simulations on galaxy evolution with different resolutions, cosmological volumes and subgrid models. They were run using a modified version of the $N$-Body Tree-PM smoothed particle hydrodynamics (SPH) code \textsc{gadget-3} \citep[][]{Springel2005}. Subgrid models are used in the simulations to model unresolved (sub-kpc) physical processes relevant to galaxy formation and evolution. As described in detail by \citet{Schaye2015}, these include radiative gas cooling, reionisation, star formation, stellar evolution and chemical enrichment, energy feedback from star formation, and the growth of supermassive black holes (BHs) and Active Galactic Nuclei (AGN) feedback. Their free parameters were calibrated so that the simulation reproduces a well-defined set of observations, specifically the $z=0.1$ stellar mass function, the $z\approx 0$ size-mass relation of star-forming galaxies and the BH-stellar mass relation (see \citealt{Crain_2015} for details). {\sc Eagle} adopts the following cosmological parameters: $\Omega_\mathrm{m}$=0.307, $\Omega_\mathrm{\Lambda}$=0.693, $\Omega_\mathrm{b}$=0.04825 and $H_\mathrm{0}$=67.77 $\mathrm{km s^{-1} Mpc^{-1}}$  \citep[][]{Schaye2015}. 

In this paper we use the largest simulation of the {\sc Eagle} suite, named as Ref-L100N1504 in \citet{Schaye2015}, which models a box of side length 100 comoving Mpc, and an initial number of particles of $2\times 1504^3$. The dark matter (DM) and initial baryon masses are $9.7\times 10^6\,\rm M_{\odot}$ and $1.81\times 10^6\,\rm M_{\odot}$, respectively. The gravitational softening length of the Ref-L100N1504 run we use here is $0.7$ proper kpc in the redshift range of interest, $z=0-2$. 

The {\sc Eagle} Ref-L100N1504 simulation has 29 snapshot outputs between redshift 20 and 0. Galaxy merger trees were created using the D-Trees algorithm \citep[][]{Jiang2014,Qu2017} to link galaxies between snapshots. We retrieve properties such as stellar masses, SFRs, halo masses, galaxy positions, velocities, and optical colours from the public {\sc Eagle} database \citep{McAlpine2016}. In addition to snapshots, {\sc Eagle} also produced $400$ lean outputs, dubbed ``snipshots'', between redshifts 20 and 0. The latter allows us to explore short timescale changes in galaxy properties, and is key to tracing the orbit of satellite galaxies.

One of the drawbacks of {\sc Eagle} is the limited volume, which leads to a poor sampling of the galaxy cluster regime. To remedy this, galaxy cluster zoom simulations were produced as part of the {\sc Hydrangea}/C-{\sc Eagle} project \citep{Bahe2017, Barnes2017a}, with the same code as {\sc EAGLE} and a closely related physical mode.

The re-simulated structures of C-{\sc Eagle} were selected from the MAssive ClusterS and Intercluster Structures (MACSIS) project \citep[][]{Barnes2017}. The MACSIS project is a dark matter-only simulation with the same cosmology of {\sc Eagle} but in a much greater cosmological volume of (3200 comoving Mpc)$^{3}$. From the MACSIS outputs at $z$=0, dark matter haloes with $M_\mathrm{200c}$ = 10$^{14} - 10^{15.4}$~$\rm M_{\odot}$ ($c$ represents comoving) were selected. Here, $M_\mathrm{200c}$ is the total mass within a sphere of radius $r_\mathrm{200c}$, centred on the gravitational potential minimum of the cluster, within which the average density equals 200 times the critical density. In addition to the mass selection, massive haloes with other massive haloes located within 30 physical Mpc (pMpc), or 20 $r_\mathrm{200c}$ were excluded to ensure the zoom-in re-simulated volume is centred on the massive halo density peak. In total, 30 clusters were simulated as part of the C-{\sc Eagle}/{\sc Hydrangea} sample (hereafter we refer to the ensemble of resimulated clusters as C-{\sc Eagle}). C-{\sc Eagle} has outputs for 30 snapshots from redshift 0 to 14. Key properties of the C-{\sc Eagle} clusters, such as $M_\mathrm{200c}$ and $r_\mathrm{200c}$ are listed in \citet[]{Bahe2017} and \citet[]{Barnes2017a}. 

C-{\sc Eagle} adopted nearly the same subgrid model as the reference {\sc Eagle} run described above, with the exception of two parameters in the AGN feedback model that control the temperature to which gas particles are heated once directly affected by AGN and the viscosity of the accretion disk. Firstly, in the Ref-L100N1504 model, the former is $\Delta\,T=10^{8.5}\,\rm K$, while in C-{\sc Eagle}, this value is changed to $\Delta\,T=10^{9}\,\rm K$, making AGN feedback more stochastic but also more powerful. Secondly, the viscosity parameter of the accretion disk, $C_{\rm visc}$, was changed from $2\pi$ in {\sc Eagle} to $2\pi\times 10^2$ in C-{\sc Eagle}, decreasing the efficiency of gas accretion onto the BH. These changes lead to a better fit to the observed gas fractions of groups (see \citealt{Schaye2015} for further details), although the difference on cluster scales turns out to be negligible \citep{Barnes2017a}.

In post-processing, particle-level stellar luminosities were calculated for both \textsc{Eagle} and \textsc{C-Eagle}. The former adopted the BC03 \citep[][]{BC2003} simple stellar populations (see \citealt{Trayford2015}), while C-{\sc Eagle} used the empirical stellar spectra library from \citet[see \citealt{Negri2022} for details]{Vazdekis2016}. This difference has a negligible impact on our results.

Both {\sc Eagle} and C-{\sc Eagle} DM halos are identified first using the ``Friends-of-Friends'' (FoF; \citealt{Davis1985,Lacey1994}) method, where DM particles are linked if their separation is less than 20 per cent of the average inter-particle distance. Baryonic particles are assigned to the FoF halo, if any, of their nearest DM particle \citep[][]{Schaye2015}. {\sc Eagle} and C-{\sc Eagle} then employs {\sc SUBFIND} (\citealt{Springel2001, Dolag2009}) to identify self-bound overdensities of particles within halos (i.e. substructures), that can be associated with galaxies in the simulations. The subhalo with the deepest gravitational potential within the FoF halo is classified as the ``central'', while the remaining substructures are flagged as ``satellites''. 

From the {\sc Eagle} and C-{\sc Eagle} simulation data catalogue, we select satellite galaxies as our focus is on environment-driven processes. In C-{\sc Eagle}, we only select galaxies that belong to the central cluster (rather than halos around the main cluster), which ensures that we are restricting the analysis to the high resolution regions. 
We also select only satellites that have at least $20$ SF gas particles (those with a $\rm SFR>0$) to have sufficient particles to calculate the SFR and the half-SFR radius. A test of the impact of this cut is presented in Appendix.~\ref{Esec:sfrnum}. Finally, we also select galaxies with log$_{10}$(sSFR/yr$^{-1}$) $\ge$ $-11.25$ to mimic the SF galaxy selection in \citet{Wang2022}.
In this paper we use C-{\sc Eagle} to analyse only the $z=0$ satellite galaxy population in massive clusters. 
We show in $\S$~\ref{Esec:compareEAGLEwithSAMI} that in the galaxy cluster regime, {\sc Eagle} and C-{\sc Eagle} predict similar $C$-index distributions. Considering that there are some structural differences between the galaxies in these two simulations due to the different AGN feedback parameters adopted, we decide to trace the history of galaxies and analyse the $C$-index evolution using {\sc Eagle} only.

Our data sample from the {\sc Eagle} simulation covers groups with halo masses in the range $M_\mathrm{200}$ = 10$^{11.5} - 10^{14.6}$ $M_{\odot}$, among which 8 groups have $M_\mathrm{200}$ > $10^{14} M_{\odot}$. By adding in the C-{\sc Eagle} simulation, we include a more extreme environment of 30 clusters with halo masses in the range $M_\mathrm{200}$ = 10$^{14} - 10^{15.4}$ $M_{\odot}$. Because C-{\sc Eagle} and {\sc Eagle} have different subgrid model parameters (see above), we keep them separate in the analysis.
Therefore, based on halo mass, galaxy environments are separately between Low-Mass Groups (LMGs, $M_{200}$ $\leq 10^{12.5} M_{\odot}$), Intermediate-Mass Groups (IMGs, $10^{12.5} M_{\odot}< M_{200} \leq 10^{13.5} M_{\odot}$), High-Mass Groups (HMGs, $M_{200}$ > $10^{13.5} M_{\odot}$) - the latter three are only for {\sc Eagle} - and clusters ($M_{200}$ > $10^{14} M_{\odot}$) -  from C-{\sc Eagle} only. After applying the sample selection, the numbers of satellite galaxies in different halo mass intervals are shown in Table~\ref{tab:number_halo}.

\subsubsection{Galaxy and environment properties}\label{Esec:orbit}\label{Esec:time_metric}

We aim to understand how the SF concentration can be used as a proxy of environmental quenching. 
With that aim, we measure the SF concentration index, $C$-index, in {\sc Eagle} galaxies in a similar fashion as what has been done in SAMI, { $r_\mathrm{50,SFR}$ is used as a proxy for $r_\mathrm{50,H\alpha}$ in }\citet{Schaefer2017}, \citet{schaefer2019}, and \citet{Wang2022}. We define the $C-{\rm index}$ as 

\begin{equation}
    C-{\rm index} = {\rm log}_{10}\left(\frac{r_\mathrm{50,SFR}}{r_\mathrm{50,rband}}\right),
\end{equation}

\noindent where $r_\mathrm{50,SFR}$ is the projected 3D radius containing half of the instantaneous SFR of a galaxy, and $r_\mathrm{50,rband}$ is the 3D radius containing half of the unattenuated r-band luminosity of a galaxy.
To match the simulation with the observation data, and prior to calculating the radii above, we apply an aperture to select particles in {\sc Eagle}/C-{\sc Eagle} galaxies. {Firstly, we align the galaxy with the stellar angular momentum vector, and calculate the radial distance of a particle to the centre in the X-Y plane (r), and distance to the midplane in the Z axis (h).} Then, we choose that aperture to mimic the SAMI apertures, which on average cover 1.4 $r_\mathrm{e}$ of each galaxy, with $r_\mathrm{e}$ being the galaxy effective radius. We test different ways of calculating $C-{\rm index}$, including, i) particles within 1.4 $r_\mathrm{50}$ along and 0.5 $r_\mathrm{50}$ from the midplane of the disk; ii) a fixed spherical aperture of 1.4 $r_\mathrm{50}$; and iii) a fixed spherical aperture of $100$kpc. There are only minor changes in the $C-{\rm index}$ and no changes in the trends reported later (see Appendix~\ref{Esec:cindex_compare} for details). Thus, in the measurement of the $C-$index, we use the first method, which includes particles within 1.4 $r_\mathrm{50}$ along and 0.5 $r_\mathrm{50}$ from the midplane of the disk. Here, $r_\mathrm{50}$ is the 3D half-stellar mass radius of the galaxy calculated with all the stellar particles that belong to the subhalo, and the midplane of the disk is the plane perpendicular to the stellar angular momentum, calculated including all stellar particles in the subhalo.

For \textsc{Eagle}, we make use of the finely-spaced snipshot outputs in order to obtain temporally precise information of the infall and quenching of satellite galaxies in a variety of environments. This is based on the orbital catalogue of \citet{Wright2022}, who employ the open-source O{\scriptsize RB}W{\scriptsize EAVER} code  \citep[][]{Poulton2020} to characterise the orbits of (sub)haloes in the simulation based on detailed merger tree information. Importantly for our purposes, O{\scriptsize RB}W{\scriptsize EAVER} identifies the turning point in satellite group-centric radial velocity to recognise the peri- and apoapses of satellite orbits. Throughout this paper, we frequently use this information to quantify $N$-pericentre (the number of pericentric passages a satellite has experienced). If $N$-pericentre is 0, galaxies have not yet passed the host halo centre. With the orbit defined, $R_{\rm closest-approach}$ is the radius characterising the closest approach of a satellite to its host and $R_{\rm current}$ is the current relative distance. These parameters are summarised at the top of Table \ref{tab:eagle_parameter}. 

In addition to orbital properties, we find the lookback times (LBTs) at which a satellite galaxy was last a central, $t_\mathrm{last-central}$, and when it joined their current host halo,  $t_\mathrm{sat}$. Note that for galaxies that have had pre-processing (i.e. they have been a satellite of another host before joining their current host), $t_\mathrm{last-central} > t_\mathrm{sat}$.
In $\S$~\ref{Esec:quenchinghighz}, we study how sSFR and $C$-index evolve with time after $t_\mathrm{last-central}$. With this aim, we calculate $\delta t_\mathrm{last-central}=t_\mathrm{last-central}-\rm LBT(out)$, where $\rm LBT(out)$ is the lookback time of that particular output. We also calculate quenching timescales ($t_\mathrm{quench}$) based on how galaxies move in the sSFR-stellar mass plane in a similar fashion as \citet[]{Wright_2019}. We first define the sSFR main sequence (MS) as log$_{10}$(sSFR/yr$^{-1}$) = $-10$ + 0.5$z$ following \citet[][]{Furlong2015}. {The distance to sSFR MS is defined as $\Delta$MS (in log space, so that $\Delta\,\rm MS=0$ means exactly on the MS) to help analysing changes in sSFR at different redshift.} We define quenched galaxies at $z=0-2$ with $\rm log_{10}(sSFR/yr^{-1}) < -11 \rm + 0.5z$ (i.e. an order of magnitude below MS), then trace back the last snapshot their $\rm log_{10}(sSFR/yr^{-1})> -10.3 + 0.5z$ (i.e. the last time that galaxy was $0.3$~dex below MS). The exact time these limits are crossed cannot be determined exactly from the simulation snapshots, so we perturb the lookback times above by a random value between $0$ and the duration of the snapshot, in the same way as was done in \citet{Wright_2019}. Note that this population of quenched galaxies has little overlap with the galaxies selected above based on the $\rm log_{10}(sSFR/yr^{-1}) > -11.25$ above, but our intention in calculating $t_\mathrm{quench}$ is to understand the connection between satellite galaxies that are yet to quench with those that already quenched in $\S$~\ref{Esec:quenchinghighz}. 

Additionally, we analyse the gas depletion timescale ($t_{\rm dep}$) of satellite galaxies as introduced by \citet[]{Wright2022}. {To derive $t_{\rm dep}$, the ``BaryMP'' fitting method outlined in \cite{Stevens2014} is applied to determine the radius at which at which the baryon profile transitions to a radial dependence $\propto r^{-2}$, R$_\mathrm{BMP}$. Such exponent is expected for an isothermal diffuse hot gas halo, or a dark matter halo in virial equilibrium. Hence a good guess for when the baryon content is dominated by halo material rather than ISM. The ISM of a galaxy is then defined by all cool gas particles (i.e. those with a temperature T < 5 × 10$^{4}$ K or a $\rm SFR > 0$) internal to R$_\mathrm{BMP}$. The gas inflow/outflow rates are calculated by comparing the ISM reservoir of two {\sc Eagle} snipshots and normalised by the time interval between the two snipshots with the Lagrangian technique. } Lastly, $t_{\rm dep}$ is defined by the time it would take a galaxy to exhaust its current gas reservoir if its inflow ($\dot{M}_\mathrm{in}$), outflow ($\dot{M}_\mathrm{out}$), and SFRs ($\dot{M}_\mathrm{\star}$) at $t$ were constant:
\begin{gather}
    t_\mathrm{dep} = \frac{M_\mathrm{ISM}}{\dot{M}_\mathrm{in}-\dot{M}_\mathrm{out}-\dot{M}_\mathrm{\star}},
	\label{eq:tdep}
\end{gather}
\noindent where $M_{\rm ISM}$ is the ISM mass at $t$. Based on $t_{\rm dep}$, galaxies are classified between those: (a) net growing or maintaining their ISM reservoir, $t_\mathrm{dep}$ $\ge$ 10Gyr; (b) slowly depleting their ISM reservoir, $1.5$~Gyr$<t_\mathrm{dep}< 10$~Gyr; (c) rapidly depleting their ISM reservoir, $t_\mathrm{dep}\le 1.5$~Gyr; (d) quenched, $\rm log_{10}(sSFR/yr^{-1})\le -11+0.5z$ (see \citealt{Wright2022} for more details). The latter sSFR criterion is used throughout the paper to define quenched galaxies, which is slightly lower than the threshold applied in \citet{Wright2022}, so there are several satellite galaxies we include but that are excluded in \citet[]{Wright2022}. 

A summary of time metrics used in our study is listed in Table~\ref{tab:eagle_parameter}.

\begin{figure*}
	\includegraphics[width=\textwidth]{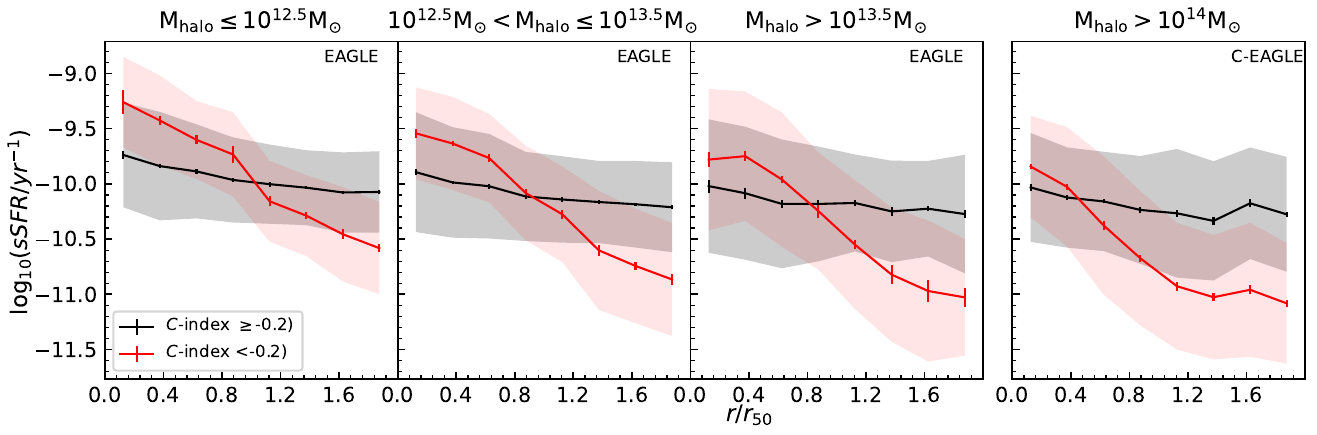}
    \caption{sSFR radial profiles of satellite galaxies in the four halo mass intervals of Table~\ref{tab:number_halo}, as labelled at the top of each panel. Profiles are shown separately for regular (black) and SF-concentrated (red). Lines, errorbars and shaded regions show the medians, error on the medians, and 1 $\sigma$ scatter, respectively. The figure shows that SF concentration is measuring both a lack of star formation in the outskirts and an excess of star formation in the centres of concentrated galaxies. Both these features can be interpreted as environmental effects.} 
    \label{fig:ssfr_rprofile}
\end{figure*}

\begin{figure*}
	\includegraphics[width=\textwidth]{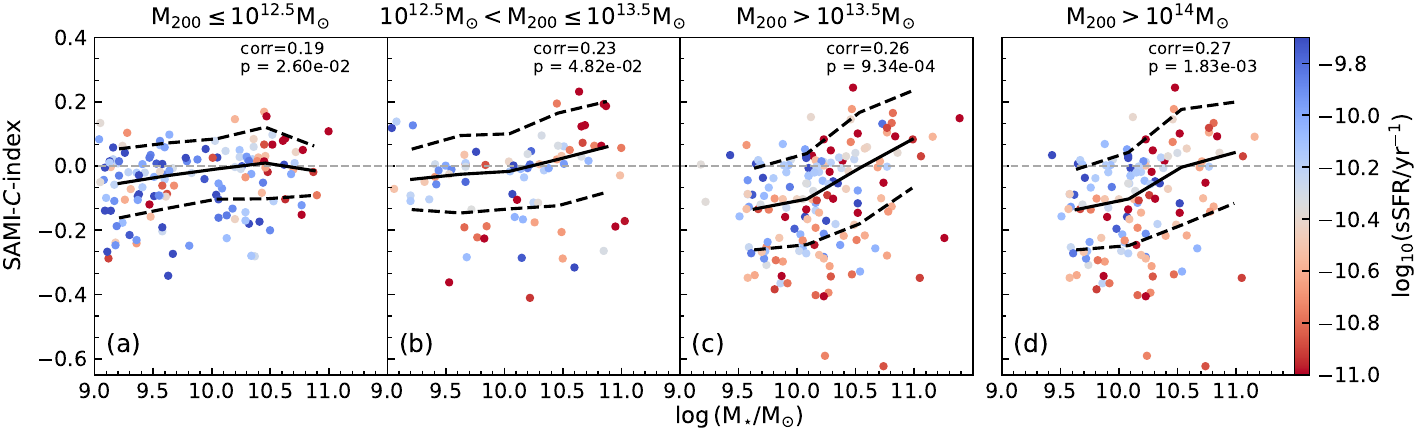}
	\includegraphics[width=\textwidth]{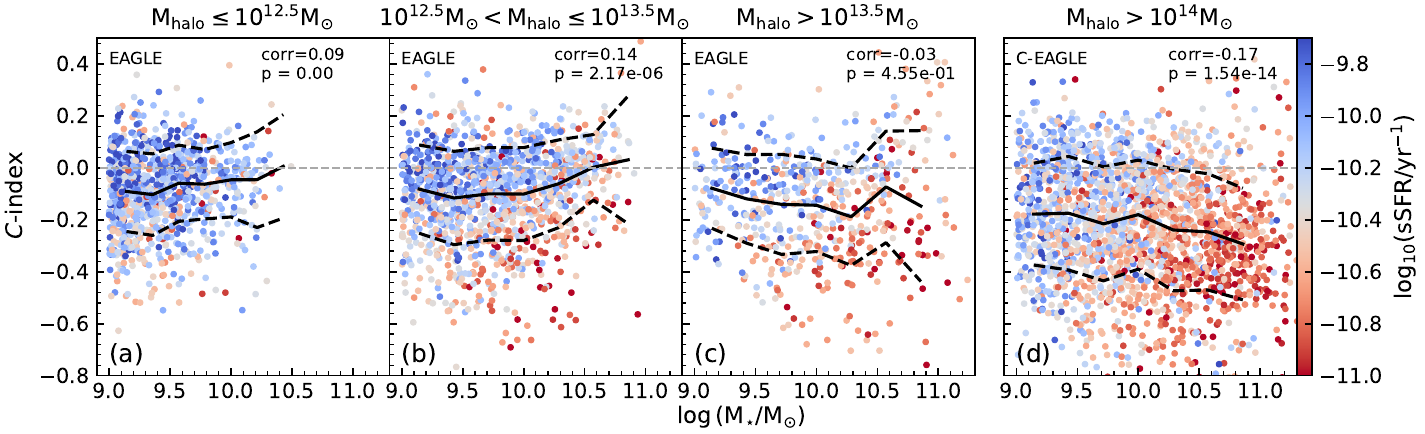}
    \caption{$C$-index as a function of stellar mass in four different halo mass intervals, as labelled at the top of each panel, colour$-$coded by sSFR. SAMI is shown in the top panels, while {\sc Eagle} and C-{\sc Eagle} are shown in the bottom panels. {\sc Eagle} is shown in the (a), (b), (c) panels, and C-{\sc Eagle} in panel (d).  Medians and the  1~$\sigma$ scatter are shown with solid and dashed lines, respectively. Also shown in each panel as the Spearman correlation coefficient and $p$-value. Galaxies in denser regions tend to have lower $C$-index in both observations and simulations. }
    \label{fig:Cvm_sami}
\end{figure*}

\subsection{The SAMI Galaxy Survey}

The SAMI Galaxy Survey \citep[][]{Croom2012} is an IFS project, using the 3.9-m Anglo-Australian Telescope (AAT), the SAMI top end has a 1-degree diameter field of view using 13 optical fibre bundles (hexabundles; \citealt{Bland2011,Bryant2011,Bryant2014}). Each bundle combines 61 optical fibres covering a circular field of view with a 15$''$ diameter on the sky. These optical fibres feed into the AAOmega spectrograph \citep[][]{Sharp2006}. The raw telescope data is reduced into two cubes using the 2dfDR pipeline (\citealt{AST2015}), together with a custom python pipeline for the later stages of reduction \citep[][]{Sharp2015}. The blue cubes cover a wavelength range of 3700$-$5700 \AA $ $ with a spectral resolution of R = 1812 ($\sigma$ = 70 km \  s$^{-1}$), and the red cubes cover a wavelength range of 6250$-$7350 \AA $ $ with a spectral resolution of R = 4263 ($\sigma$ = 30 km \ s$^{-1}$) at their respective central wavelengths \citep[][]{Sande_2017}. SAMI data is targeted based on cuts in the redshift-stellar mass plane \citep[]{Bryant2015}, in the local Universe at $z < 0.1$. This paper uses the SAMI third and final data release (DR3; \citealp{Croom_2021}), together with value-added products such as emission line fits and stellar population measurements. We select SAMI galaxies with the same mass range, and halo mass ranges as the {\sc Eagle} simulation data.
SAMI adopts $\Omega_\mathrm{m}$=0.3, $\Omega_\mathrm{\Lambda}$=0.7 and $H_\mathrm{0}$=70 $\mathrm{km s^{-1} Mpc^{-1}}$ as cosmological parameters. {As these cosmology parameters are different from the {\sc Eagle} simulation, masses, SFRs and sizes have been corrected to $H_\mathrm{0}$=67.77 $\mathrm{km s^{-1} Mpc^{-1}}$, which is the value adopted in {\sc Eagle}.}

The SAMI observation data covers environments of GAMA groups \citep[]{Driver2011,Bryant2015,Robotham2011} and 8 clusters \citep[]{Owers2017}. For SAMI galaxies, we use the sample selection of \citet{Schaefer2017} and \citet{Wang2022}: 

\begin{enumerate}
\item Effective radii ($r_\mathrm{e}$) $\le$ 15$^{\prime\prime}$, to reduce the effect of SAMI aperture effect on measuring the spatial distribution of star formation; 
\item Ellipticity values $\le$ 0.7, to remove edge on galaxies as they hide spatial information and will increase the uncertainty when calculating spatial star-formation properties when assuming galaxies have elliptical isophotes;  
\item Seeing/$r_\mathrm{e} <$ 0.75 and seeing $<$ 4$''$, to reduce the effect of beam smearing on small galaxies; 
\item H$\alpha$ equivalent widths (EW$_\mathrm{H\alpha}$, spatially integrated over the SAMI cube) greater than 1\,\AA, and log$_{10}$(sSFR/yr$^{-1}$) $\ge$ $-11.25$ to select SF galaxies. 
\end{enumerate}

\citet{Schaefer2017,schaefer2019,Wang2022} calculated the star-formation concentration index (SAMI-$C$-index) as $\log_{10}(r_\mathrm{50,H\alpha}/r_\mathrm{50,cont})$ in SAMI, where $r_\mathrm{50,H\alpha}$ and $r_\mathrm{50,cont}$ are the half-light radii of the dust corrected H$\alpha$ and the r-band continuum light. The lower the SAMI-$C$-index the more centrally concentrated the SF is in a galaxy. To deduce $r_{50}$ of the H$\alpha$ and r-band continuum, the flux curve of growth is calculated assuming that galaxies are idealised thin discs and their observed ellipticity is due to their inclination. The uncertainty on $r_\mathrm{50,H\alpha}/r_\mathrm{50,cont}$ is calculated by adding a random error based on Gaussian distributions on the measured values of the H$\alpha$, H$\beta$ fluxes, ellipticities and position angles, resulting an average an $0.02$~dex error (more details in \citealt{schaefer2019}). To further quantify the distribution of regular galaxies and galaxies with concentrated SF regions (SF-concentrated galaxies) in different halo masses, \citet[]{Wang2022} separated the SAMI galaxy sample into two $C$-index intervals: SAMI-$C$-index $\ge$ $-0.2$ and SAMI-$C$-index $<$ $-0.2$. For SAMI SF galaxies with LINER/AGN spaxels, we apply a LINER/AGN correction as described in \citet[]{Wang2022}, so only H$\alpha$ from SF is included in the analysis. The SAMI sample in halo mass intervals is shown in Table~\ref{tab:number_halo}. Note, SAMI cluster galaxies are selected from the HMGs, but explicitly in haloes with $M_{200}$ > $10^{14} M_{\odot}$. 

\begin{figure*}
	\includegraphics[width=\textwidth]{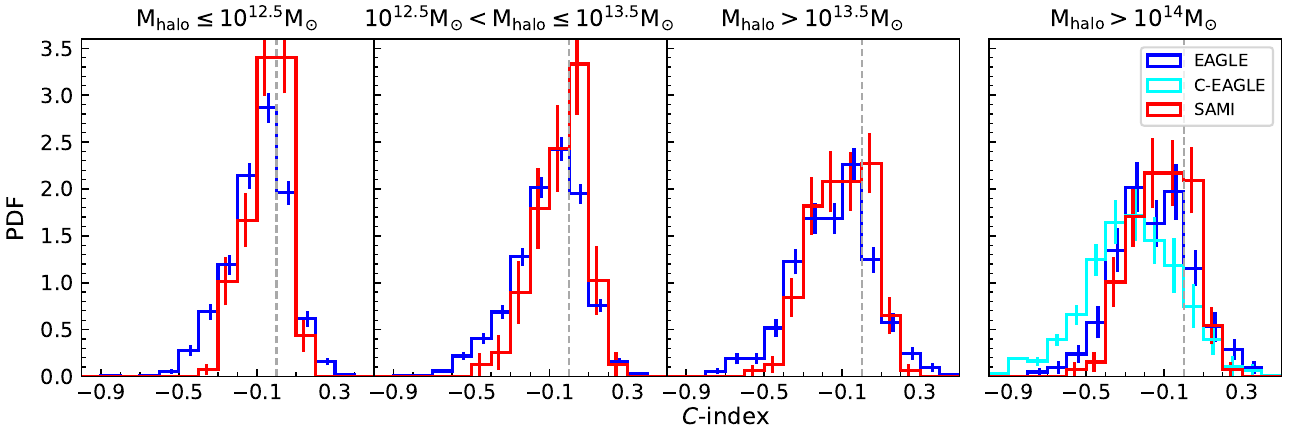}
    \caption{Normalized histogram of {\sc Eagle}/C-{\sc Eagle} $C$-index at $z$=0 in four halo mass intervals. Blue for {\sc Eagle} galaxies, cyan for C-{\sc Eagle} galaxies and red for SAMI galaxies with bootstrap uncertainties. Both {\sc Eagle} and SAMI galaxies have a larger fraction of low $C$-index in higher halo mass. In groups, the mode $C$-index for SAMI galaxies is around 0.1, slightly greater than the {\sc Eagle} mode $C$-index around 0. There is a more significant difference in C-{\sc Eagle} $C$-index distribution in clusters compared to SAMI galaxies. However, they show that galaxies in denser regions tend to have more low $C$-index galaxies.  }
    \label{fig:c-hist}
\end{figure*}

\section{star formation concentration as a tracer of environmental quenching in the local Universe} \label{Esec:z0}

In this section we focus on key relations between $C$-index and other galaxy properties at $z=0$, and on comparing {\sc Eagle}/C-{\sc Eagle} with SAMI. For that we focus on satellite galaxies in the simulations at $z=0$.

\subsection{What the $C$-index is measuring?}

A basic question is what is $C$-index measuring, and whether it is a good proxy for environmental quenching. To address this question, Fig.~\ref{fig:ssfr_rprofile} shows radial profiles of sSFR for galaxies in the halo mass bins of Table~\ref{tab:number_halo} and separating galaxies with concentrated and non-concentrated SF. We use the threshold $C$-index$=-0.2$ to distinguish between these two categories, following \citet{Wang2022}. The figure shows that in all environments, SF-concentrated galaxies (i.e. those with $C-{\rm index} < -0.2$) have higher sSFR at the centre and lower sSFR in the outskirts compared to their non-SF concentrated counterparts, with the transition happening at $\approx 0.8-1\, r_{\rm 50}$, depending on the halo mass range. Three important trends take place as we move to higher halo masses: (i) the central sSFR of concentrated galaxies gets closer to the non-concentrated ones; (ii) the deficiency of sSFR in the outskirts becomes more pronounced in concentrated galaxies; (iii) and the transition radius moves towards lower radii. All these trends can be interpreted as environmental quenching in action: the outskirts of galaxies host the least bound, lower density gas that is more susceptible to RPS, and the higher intra-halo medium density of higher mass halos allows RPS to act deeper into the galaxy's potential. The higher sSFR in the centre could be interpreted as the gas in the internal parts starbursting.  Note, the high sSFR in the centre is also by selection given the SF concentration measurement. The median stellar mass of the concentrated vs non-concentrated samples are within $0.1$~dex from each other, and this sSFR offset in the centres of concentrated galaxies is not driven by stellar mass differences. \citet{Troncoso2020} analysed satellite galaxies in {\sc Eagle} in halos with masses $\ge 10^{14}\,\rm M_{\odot}$ and found that they had higher ISM pressure than main sequence galaxies of the same stellar mass, which leads to an enhanced star formation efficiency. This partially explains why concentrated galaxies have a higher sSFR towards the centre.

\subsection{Comparing {\sc Eagle}/C-{\sc Eagle} with SAMI} \label{Esec:compareEAGLEwithSAMI}

\begin{figure*}
	\includegraphics[width=15cm]{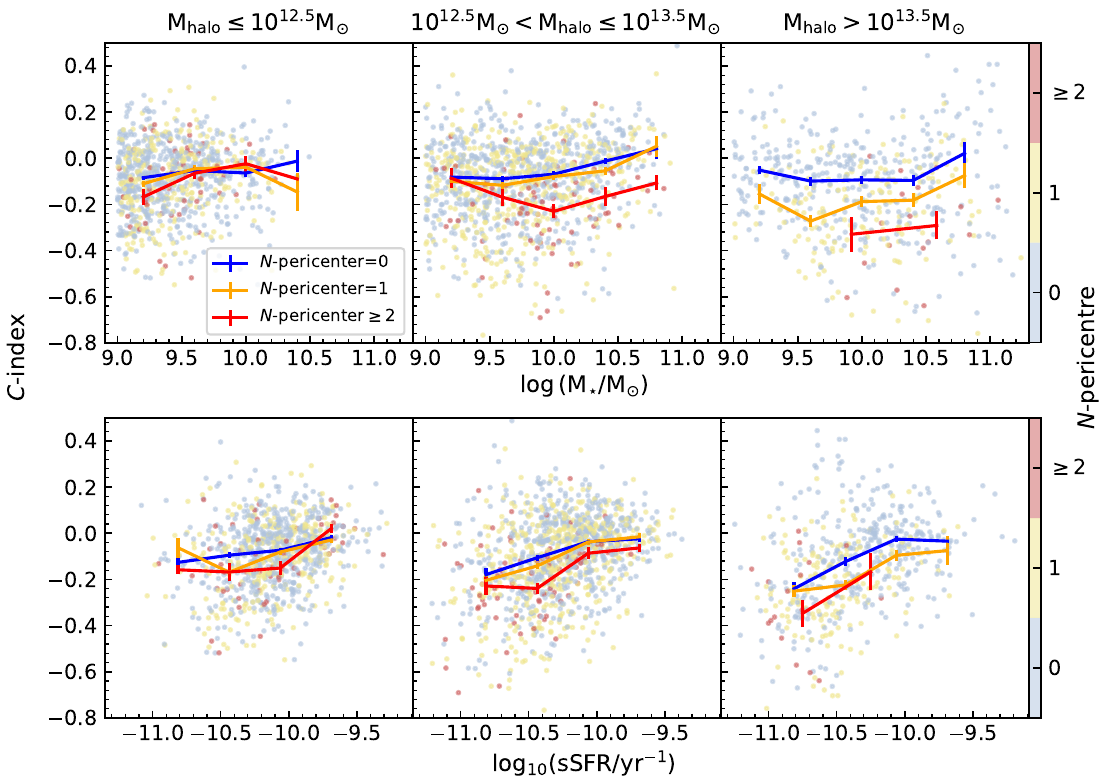}
    \caption{$C$-index as a function of stellar mass (top panel) and sSFR (bottom panel) for satellites at $z$=0 in {\sc Eagle} in different host halo mass bins, as labelled. Each galaxy is colour coded by the number of pericentric passages ($N$-pericentre) they have had in the current host halo. Solid lines with errorbars show the medians and errors on the medians for the subsamples that have had $N\rm -pericentre\equiv 0$ (blue), $\equiv 1$ (yellow) and $\ge 2$ (red). Generally, satellite galaxies that have had one or more peri-centre passages have a lower $C$-index, even at fixed sSFR. }
    \label{fig:CvsMnperi}
\end{figure*}

Fig.~\ref{fig:Cvm_sami} shows $C$-index as a function of $M_{\star}$ for satellite galaxies in $4$ bins of halo mass in SAMI and {\sc Eagle}/C-{\sc Eagle}, colour-coded by sSFR. 
The number of galaxies in each halo mass interval is shown in Table~\ref{tab:number_halo}. We remind the reader that galaxies with a $C$-index $<0$ have H$\alpha$ emission that is more compact than their r-band continuum. 

\citet[]{Wang2022} found that denser environments tend to have a more extended range of $C$-index values and a larger tail towards low values. \citet[]{Wang2022} also found higher halo masses tend to have higher fractions of SF-concentrated galaxies than low halo masses. Similarly to the SAMI galaxies in LMGs, galaxies with stellar masses of 10$^{9-11.5}M_{\odot}$ form a locus around $C$-index=0, while more galaxies in HMGs show concentrated star formation. 

In {\sc Eagle}/C-{\sc Eagle}, 
the galaxy stellar mass range of satellite galaxies increases  with the halo mass. Only galaxies with $M_{\star}$ < 10$^{10.5}$ $M_{\odot}$ are found in LMGs. 
Similar to the observational data, galaxies in higher density environments tend to have a large range of $C$-index and more low-$C$-index values are seen in HMGs. We can see a wider range of $C$-index for C-{\sc Eagle} clusters than for {\sc Eagle} HMGs, which simply reflects the higher density environments simulated by C-{\sc eagle}. 
We note that in SAMI there is a weak positive correlation between $C$-index and stellar mass, which is the opposite of what is seen in the simulations. This is not necessarily a concern because (i) the Spearman correlation values in both cases are low (<0.3), 
{(ii) SAMI-$C$-index is calculated using the dust corrected H$\alpha$, with more dusty galaxy centres leading to a greater dust correction of H$\alpha$ \citep[more details in][]{Schaefer2017}}. Regarding (ii), if dust effects are larger than assumed in SAMI in the centres of galaxies, then that would naturally bias the SAMI-$C$-index high. {Alternatively, we note that \citet[]{Wang2022} showed that many of the massive galaxies ($M_{\star}>10^{10.5}$ $M_{\odot}$) and high $C$-index have a central AGN or LINER component. Therefore, a difference could also be caused by {\sc Eagle}/{\sc C-EAGLE} does not sufficiently recover AGN feedback in the centre of galaxies, leaving too much star formation in their centres.}

With the SAMI observational data, the fraction of SF-concentrated galaxies (SAMI-$C$-index $<$ $-0.2$) is 10$\pm$3 per cent for LMGs, 13$\pm$4 per cent for IMGs, 29$\pm$4 per cent for HMGs and 29$\pm$4 per cent for clusters. We calculate the fraction of SF-concentrated galaxies in the {\sc Eagle} sample to be 22$\pm$2\%, 27$\pm$1\% and 39$\pm$2\% for LMGs, IMGs and HMGs, respectively (32$\pm$3\% for galaxies selected within cluster halo mass interval). The C-{\sc Eagle} sample has a fraction of 51$\pm$2\% of concentrated galaxies. These percentages are in general higher than those found in SAMI, consistent with the more pronounced low $C$-index tails seen in the simulations. Part of this discrepancy is due to the different underlying stellar mass distributions of the simulations and SAMI. To confirm this we match the stellar mass distributions of {\sc Eagle}/C-{\sc Eagle} galaxies to the SAMI mass distribution by randomly selecting {\sc Eagle}/C-{\sc Eagle} galaxies based on SAMI stellar mass histogram. We do this at the total sample level rather than in individual halo mass bins due to the small number of massive galaxies in LMGs in the simulations compared to SAMI. The matching reduces the number of {\sc Eagle} galaxies in each halo mass bin to 178, 452, 309 in LMGs, IMGs and HMGs, respectively (132 galaxies selected in the cluster halo mass regime); and to 560 galaxies in C-{\sc Eagle}. The fraction of SF-concentrated galaxies ($C$-index $<$ $-0.2$) in the stellar mass matched {\sc Eagle} sample becomes smaller except for HMGs, with now 19$\pm$2\%, 21$\pm$1\% and 44$\pm$3\% for LMGs, IMGs and HMGs, respectively (34$\pm$4\% for galaxies selected within cluster halo mass interval). The stellar mass matched C-{\sc Eagle} sample, however, has a fraction that increases slightly to 58$\pm$2\% of concentrated galaxies. {With the above comparison, we show that even though we match the stellar mass, the fraction of SF-concentrated galaxies is still higher compared to the SAMI data.}

To further quantitatively compare the $C$-index distribution of satellite galaxies in the simulations with observation, we show the probability distribution function (PDF) with bootstrap uncertainties of the {\sc Eagle}/C-{\sc Eagle} $C$-index and SAMI-$C$-index in Fig.~\ref{fig:c-hist}. The histograms of SAMI and {\sc Eagle}/{\sc C-EAGLE} have a relatively similar shape in all halo mass intervals. 
Some interesting differences arise, however. %
The simulations tend to have a more pronounced tail towards lower $C$-index values than SAMI in all halo mass bins. The mode of the {\sc Eagle}-$C$-index is $\approx 0$, slightly beneath the SAMI-$C$-index, which is $\approx 0.1$ in all environments.

For C-{\sc Eagle}, the $C$-index distribution is more offset; with C-{\sc Eagle} galaxies having on average a lower $C$-index than {\sc Eagle} and SAMI galaxies, and a more pronounced tail towards very low values of $C$-index (i.e. $\lesssim -0.5$).
This is consistent with the much higher environment densities simulated in C-{\sc Eagle}. In fact, the prediction here would be that more massive clusters are expected to host more SF-concentrated galaxies than lower mass clusters. It is important to also highlight that there are differences in some of the parameters adopted for the sub-grid physics in C-{\sc Eagle} compared to {\sc Eagle}. \citet{Schaye2015} showed that in the model adopted in C-{\sc Eagle}, galaxies tend to have lower mass central BHs compared to the reference model adopted in {\sc Eagle} at fixed stellar mass at $M_{\star}<10^{11}\,\rm M_{\odot}$, which on average can lead to weaker AGN feedback. {Besides the AGN feedback heating temperature difference, the more important change on galaxy scales is the reduced C$_{\rm visc}$ (C$_{\rm visc}$ controls the sensitivity of the black hole accretion rate to the angular momentum of the gas, \citealt{Schaye2015,RosasGuevara2015}) which leads to significantly higher centrally concentrated star formation, and stronger central bulges, than in the {\sc EAGLE} model.} These differences are however unlikely to cause the shift in $C$-index in C-{\sc Eagle} seen in Fig.~\ref{fig:c-hist}, which is dominated by galaxies of masses $\lesssim 10^{10.3}\,\rm M_{\odot}$ which are less affected by AGN feedback in both runs.

The overall reasonable agreement we find between the simulations and SAMI, and the fact that in {\sc Eagle}, the $C$-index parameter is indicating ``outside-in'' quenching, gives us confidence to use the simulations to explore how $C$-index traces quenching in satellite galaxies across cosmic time.

\begin{figure*}
	\includegraphics[width=15cm]{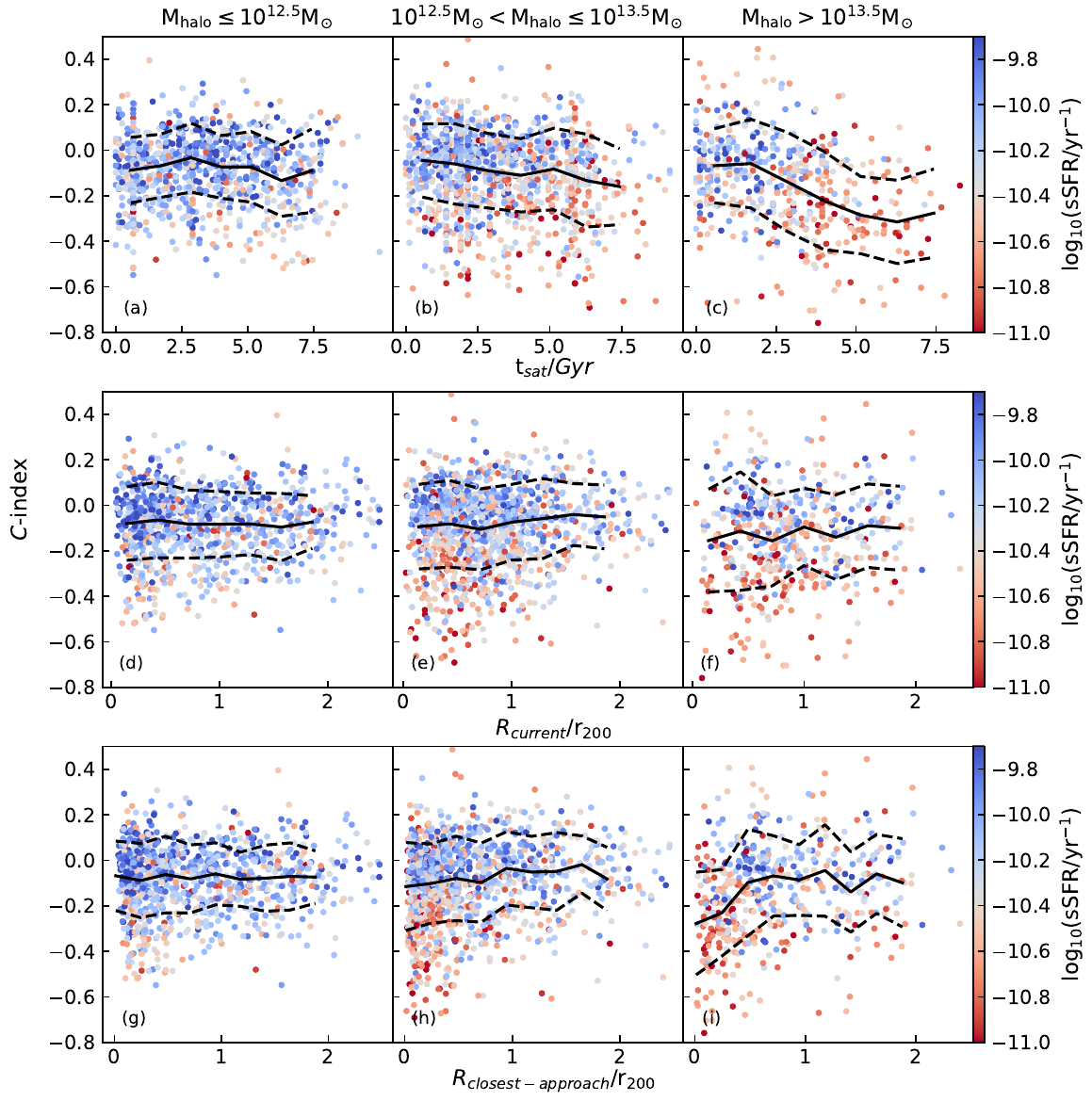}
    \caption{$C$-index of satellite galaxies at $z=0$ in {\sc Eagle} as a function $t_\mathrm{sat}$ (top panels), $R_{\rm current}$ (middle panels) and $R_{\rm closest-approach}$ (bottom panels); see Table~\ref{tab:eagle_parameter} for definitions of all these parameters;  in three different halo mass intervals, as labelled at the top. Individual galaxies are colour-coded by sSFR, as indicated by the colourbar. The radii in the middle and bottom panels are normalised by the host halo's $r_{\rm 200}$. Medians and 1 $\sigma$ percentiles are shown as solid and dashed lines, respectively. 
    Panels (a) to (c) show that a clear correlation emerges where $C$-index is lower in galaxies that have been satellites for longer, but only in HMGs. There is no correlation between the current group-centric distance of the galaxies and $C$-index based on panels (d)-to-(f). In panels (g)-to-(i), we find a correlation emerges so that $C$-index is smaller in galaxies that have been closer to the group centre in HMGs, regardless of when that happened. }
    \label{fig:Cvsdtsat}  
\end{figure*}

\begin{figure*}
	\includegraphics[width=15cm]{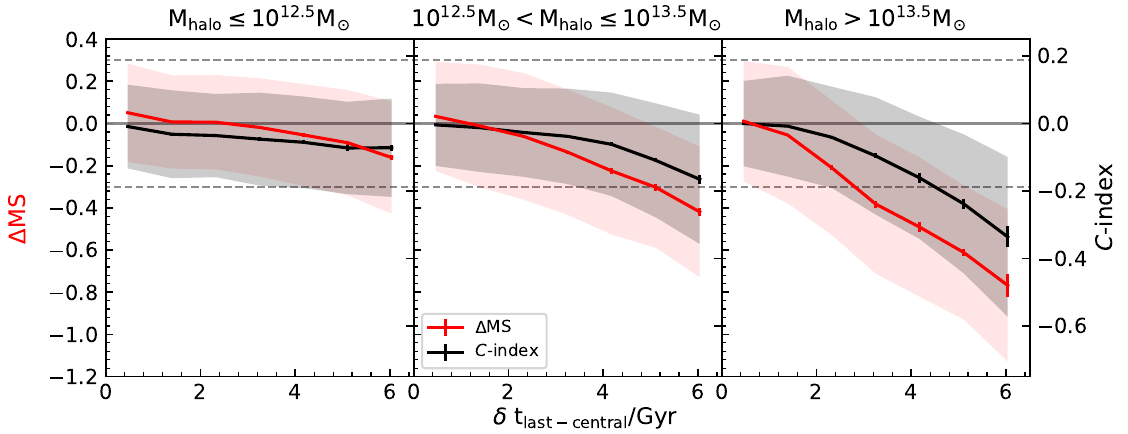}
    \caption{The time evolution of the distance to the main sequence (red; $\Delta$MS) and $C$-index (black) of $z=0$ {\sc Eagle} satellite galaxies. Note that the x-axis shows $\delta t_\mathrm{SF}$, which is the time since galaxies became satellites. The time evolution of these quantities in individual satellites is only calculated in time steps where galaxies have $\rm log_{10}(sSFR/yr^{-1})\ge -11.25 + 0.5\,z$. The horizontal gray lines in each panel highlights the MS (solid) $\pm$ 0.3~dex (dashed). Lines with errorbars show the medians and errors on the medians. The 1~$\sigma$ scatter is shown as shaded area. This figure shows that changes in $\Delta$MS are well traced by changes in $C$-index, and that these quantities in LMG's satellite galaxies are less affected at fixed time post becoming satellite than satellites in higher mass groups.}
    \label{fig:sfr_cindex_vst_s}  
\end{figure*}

\begin{figure*}
	\includegraphics[width=\textwidth]{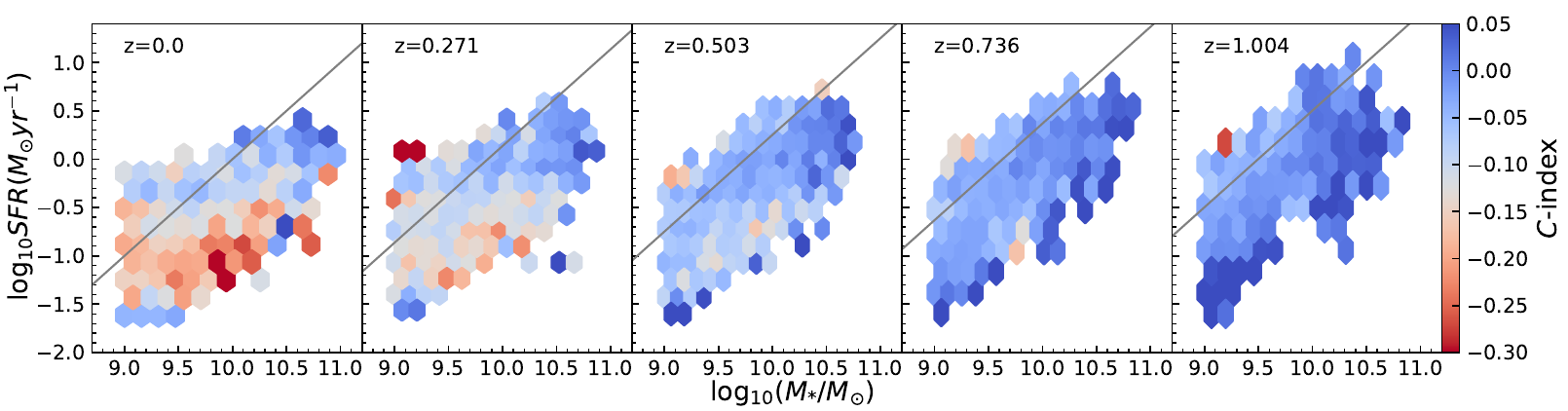}
    \caption{SFR as a function of $M_{\star}$ for {\sc Eagle} satellite galaxies selected at different redshifts, as labelled in each panel. Bins in this space are coloured by the median $C$-index of satellites in that bin. Only bins with $\ge 5$ galaxies are shown. Solid lines show the SFR MS at each redshift, which is described by $\rm log_{10}(sSFR/yr^{-1})= -10 + 0.5\,z$ in {\sc Eagle} \citep{Furlong2015}. 
    At $z=0$, there is a clear correlation between SFR and $C$-index, which becomes weaker with increasing redshift. At $z=1$, in fact, SF-extended galaxies are on average below the MS.}
    \label{fig:sfrvsm}
\end{figure*}

\subsection{The star formation concentration index as a proxy of quenching  }

We see galaxies in denser environments tend to have lower $C$-indices in both SAMI and {\sc Eagle}. In {\sc Eagle}, we can obtain extra information not available from observations, such as the number of pericentre passages for each galaxy ($N$-pericentre) in each {\sc Eagle} galaxy which shows how many times the galaxy passes through the group centre. If the $N$-pericentre is 0, galaxies have not yet passed the group centre. The top panel of Fig.~\ref{fig:CvsMnperi} shows the $C$-index versus stellar mass colour-coded by $N$-pericentre for {\sc Eagle} galaxies. We calculate the median $C$-index with bootstrap uncertainties in the different stellar mass bins for $N$-pericentre equals 0, 1 and $\ge$ 2. Note that the number of galaxies with $N$-pericentre $=0$ and $=1$ with a $\rm sSFR\ge 10^{-11.25}\,yr^{-1}$ is several hundreds in each of the halo mass bins. The number of galaxies with $N$-pericentre $\ge 2$ and $\rm sSFR\ge 10^{-11.25}\,yr^{-1}$ is smaller, with $56$, $80$ and $17$ galaxies in the LMG, IMG and HMG bins, respectively, and hence the median for that sample in HMGs is affected by low-number statistics and biased towards the galaxies that have managed to maintain SF despite $\ge$ 2 pericentre passages. Most satellites with $N$-pericentre $\ge 2$ are quenched and hence measuring $C$-index and sSFR is not possible for these galaxies.

In LMGs, $C$-index is less correlated with $N$-pericentre as galaxies will be less affected in lower-density environments. In IMGs and HMGs, there is an apparent offset for galaxies with $N$-pericentre $\ge$ 2. Therefore, in denser environments, galaxies will have less star formation in the outskirts when passing the centre multiple times. This is already clear if we only focus on the difference between galaxies with $N$-pericentre $=0$ and $=1$, with these galaxies showing a larger offset in the median $C$-index in HMGs than in IMGs. Interestingly, we see lower $C$-index even at fixed sSFR in galaxies that have had $N$-pericentre$\ge 1$ compared to those that have $N$-pericentre$=0$ (bottom panels Fig.~\ref{fig:CvsMnperi}). For the HMGs, there is already a noticeable difference between satellites with $N$-pericentre$=0$ and $N$-pericentre$=1$ at fixed sSFR, while for LMGs and IMGs, the differences are seen only between the $N$-pericentre$\ge 2$ galaxies and the rest. These results show that $C$-index and sSFR of satellite galaxies do not exactly trace each other (even though they are correlated; see Fig.~\ref{fig:Cvm_sami}), and that there is additional valuable information in $C$-index in understanding satellite galaxy quenching.

Fig.~\ref{fig:Cvsdtsat} shows $C$-index versus $t_\mathrm{sat}$ (top), the $z=0$ distance of the satellite galaxy from the centre of its host halo, $R_{\rm current}$ (middle), and the closest halo-centric distance the satellite has had,  $R_{\rm closest-approach}$ (bottom), with galaxies colour-coded by their sSFR, in the three halo mass bins labelled. We only show {\sc EAGLE} satellite galaxies at $z=0$ that have a $\rm sSFR > 10^{-11.25}\, yr^{-1}$. 
This figure shows that $C$-index correlates quite well with how long galaxies have been satellite galaxies, especially in HMGs. On the contrary, the current halo-centric position of satellites is a poor predictor of a galaxy's sSFR, $C$-index and $t_{\rm sat}$. The bottom panel of Fig.~\ref{fig:Cvsdtsat} shows that in HMGs there is a clear trend of lower $C$-index in satellites that have been at a distance $\lesssim 0.8\,\rm r_{200}$ from the group centre. 
Our results agree with previous studies, which suggest quenching is more related to the closest approach radius rather than the current position of satellite galaxies in groups and clusters \citep[][]{Oman2020}.

Figs.~\ref{fig:CvsMnperi}~and~\ref{fig:Cvsdtsat} suggest that 
$C$-index can be used not only as a measurement of quenching in action, but also as a proxy of how long star-forming satellites galaxies have been satellites.
To further investigate this, we study how $C$-index and sSFR change with time after galaxies become satellites in Fig.~\ref{fig:sfr_cindex_vst_s}. 
As the median sSFR of star-forming galaxies (a.k.a. main sequence) evolves with redshift, we use $\Delta$MS (sSFR by subtracting the {\sc Eagle} sSFR-redshift MS), for which we adopt log$_{10}$(sSFR/yr$^{-1}$) = $-10$ + 0.5$z$ \citep[][]{Furlong2015}. For the x-axis time scale, we set to 0 the time the galaxy was last identified as a central ($t_\mathrm{last-central}$), and plot evolutionary tracks until the galaxy becomes passive [log$_{10}$(sSFR/yr$^{-1}$) < -11.25]. The x-axis time is thus defined as $\delta t_\mathrm{last-central}$ = $t_\mathrm{last-central}$ - $t_\mathrm{z}$, where $t_\mathrm{z}$ is the lookback time of a snapshot.  We plot $\Delta$MS (red) and $C$-index (black) versus $\delta t_\mathrm{last-central}$ in Fig.~\ref{fig:sfr_cindex_vst_s}. 

\begin{figure*}
	\includegraphics[width=15cm]{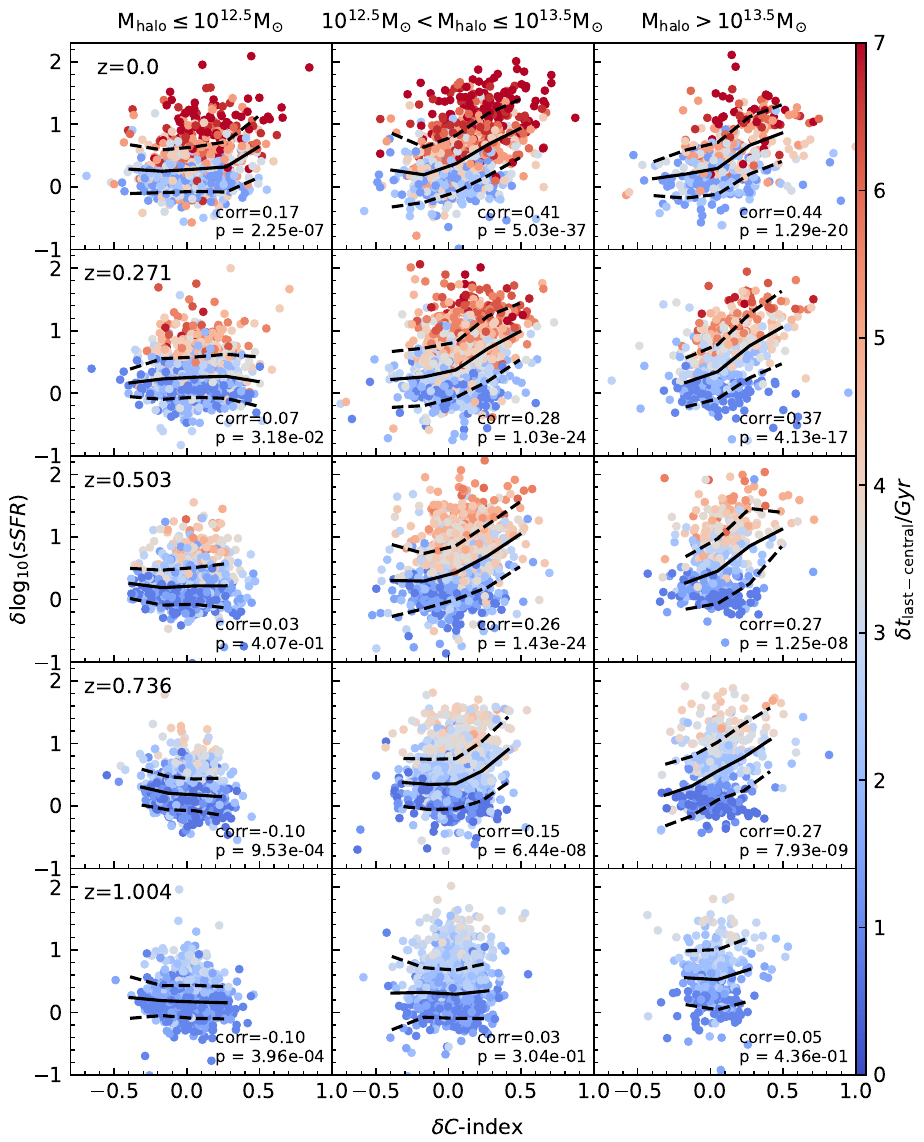}
    \caption{$\delta$sSFR versus $\delta C$-index for {\sc Eagle} galaxies colour-coded by $\delta t_\mathrm{last-central}$ at redshifts $0-1$, as labelled. $\delta$sSFR ($\delta\,C$-index) is the difference between sSFR ($C$-index) at $t_\mathrm{last-central}$ and the corresponding lookback time of the output redshift (or the last time the galaxy had a $\rm log_{10}(sSFR/yr^{-1})=-11.25$ if at the output redshift the galaxy has a lower sSFR). Greater $\delta\,C$-index means galaxies become more concentrated. The medians and 1~$\sigma$ percentile ranges are as solid and dashed lines, respectively. In each panel we also show the Spearman correlation coefficient and $p$-value. 
    At $z\le 0.5$ satellite galaxies in IMGs and HMGs exhibit a clear correlation between $\delta$sSFR versus $\delta C$-index that is not present at $z=1$.}
    \label{fig:dssfrvsdcindex}
\end{figure*}

In LMGs, both $\Delta$MS and $C$-index show a small decline with time, but this weak decline happens in tandem in both $\Delta$MS and $C$-index, even if the median sSFR remains within the scatter of the MS. When moving to higher mass halos, both $\Delta$MS and $C$-index show a steeper decline with time, and we see a systematic effect of the decline becoming steeper as we go from IMGs to HMGs. Note that in our sample satellites in IMGs take $\approx 5$~Gyrs to leave the MS, while in HMGs this takes $\approx 3$~Gyr. We remind the reader that our satellite galaxy sample is biased towards star-forming galaxies by our $z=0$ selection of $\rm log_{10}(sSFR/yr^{-1})>-11.25$. This figure shows that $C$-index is a clear proxy for quenching in $z=0$ satellite galaxies. In the next section, we explore how well this holds when we study satellite galaxies at higher redshifts.

\section{Quenching of satellite galaxies across cosmic time}\label{Esec:quenchinghighz} 

In this section we investigate how $C$-index traces quenching at different cosmic times. Here we only analyse {\sc Eagle}. 

At $z>0$, we apply the same sample selection criteria as we do at $z=0$ ($\S$~\ref{Esec:sample_selection}). This is to ensure $C$-index can be well measured. The trends discussed in this section were explored up to $z=2$ (where the number of satellites is $\approx 1/3$ of the number of $z=0$ satellites). At higher redshifts the number of satellite galaxies drops considerably and they are primarily in LMGs. We find that at $z>1$ trends are similar to those at $z=1$ and hence we limit ourselves to showing the evolution of satellites up to $z=1$ in this section and comment on how these hold at higher redshifts.

\subsection{The relationship between the SF concentration index and SF quenching across cosmic time}\label{Esec:ssfrtimeprofile}

We calculate the $C$-index parameter for all satellite galaxies up to $z=3$, using the same method as described in $\S$~\ref{Esec:methord}. We first investigate how $C$-index changes with the position of satellite galaxies in the SFR-$M_{\star}$ plane in Fig.~\ref{fig:sfrvsm}. At $z$=0, satellite galaxies with concentrated SF (redder hexbins in Fig.~\ref{fig:sfrvsm}) are preferentially located below the main sequence, which indicates that concentrated SF is associated with lower global SF.
The trend is clearly visible at $z=0.271$ and weakly at $z=0.503$. By $z=0.736 \sim 1$ it has disappeared and in fact galaxies below the main sequence appear to have {\it higher} $C$-index. Part of the change in the trend seen at $z=0$ is due to the range of $C$-index values in satellite galaxies shrinking. At $z$=0 and 0.27, $C$-index range from approx -0.8 to +0.4; at $z$=0.5, the ranges changes to $-0.5 \sim 0.3$; and at $z$=1, $C$-index covers a range of $-0.3 \sim 0.3$. {Although there appears to be a slight preference for positive values below the main sequence at $z$=1, the differences are so small (deviations of $\sim$0.05 dex around 0) that we cannot use those to argue for inside-out quenching. More careful analysis of individual sSFR profiles and their evolution as galaxies orbit within clusters would be required to assert this, and we lave that for future work.}

\begin{figure*}
	\includegraphics[width=13cm]{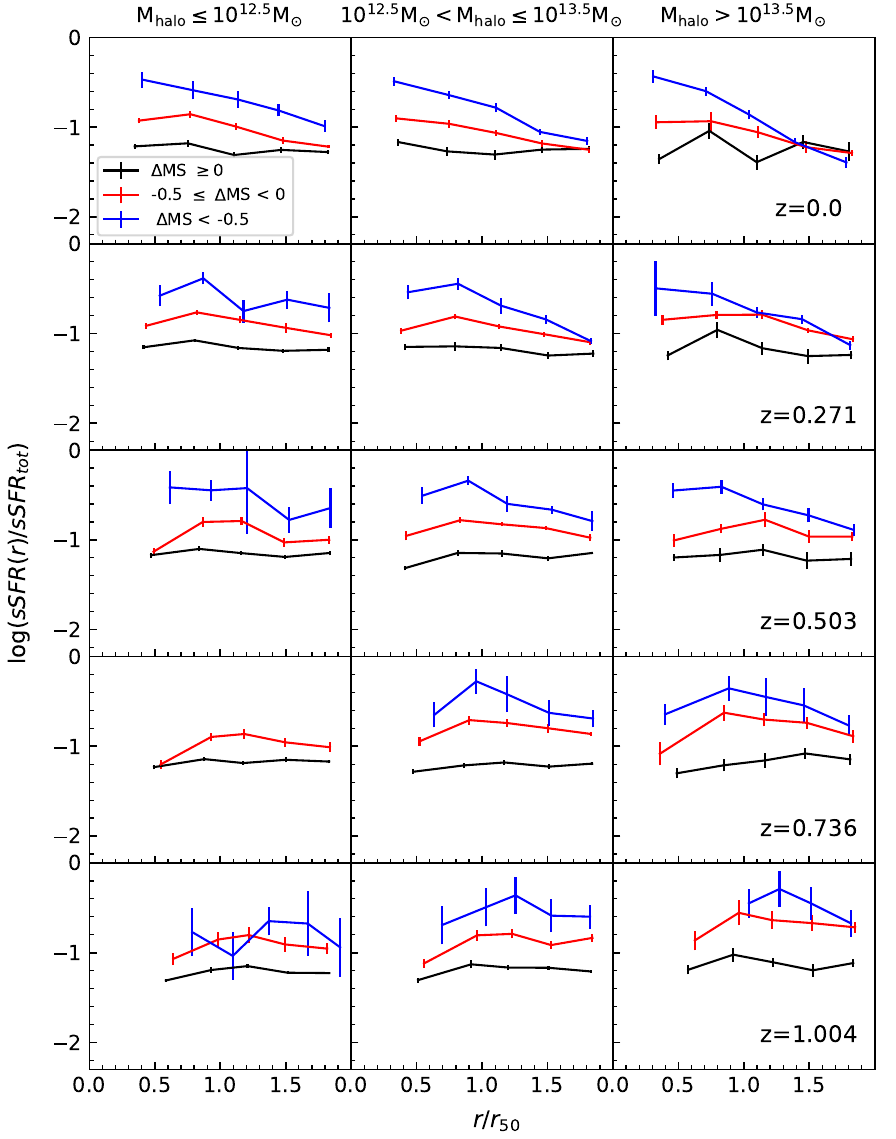}
    \caption{The sSFR radial profile normalized by total sSFR for {\sc Eagle} satellite galaxies at $z=0$ to $1$ in three halo mass bins, as labelled. We show the medians with errors on medians as solid lines with errorbars for three sSFR regimes: $\Delta\,\rm MS \ge 0$ (black); $-0.5<\Delta\,\rm MS<0$ (red); $\Delta\,\rm MS < -0.5$ (blue). We only plot medians with $\ge 10$ galaxies in each bin. {The $r/r_{50}$ bins have an equal number of galaxies for each line.} Note that at $z>0.7$ there are not enough satellites in LMGs with $\Delta\,\rm MS < -0.5$. At all redshifts, sSFR above MS (black lines) has a flat radial profile, which equates to larger $C$-index, while the sSFR profiles become steeper for galaxies below the MS, especially at $z\le 0.5$ and in IMGs and HMGs.}
    \label{fig:ssfr_profile_all}
\end{figure*}

To study how environment affects the sSFR and the $C$-index of satellite galaxies, we focus on the sSFR and $C$-index changes in the time since the galaxies first became satellites ($t_\mathrm{last-central}$). We calculate these changes between $t_\mathrm{last-central}$ and the last time a measurement of $C$-index was possible, which happens when galaxies have a $\rm log_{10}(sSFR/yr^{-1}) \ge -11.25$ 
- we refer to this time as last-measurement in equations below. The changes in sSFR and $C$-index are defined in the following way: 

\begin{equation}
    \rm \delta sSFR = log_{10}\left(\frac{sSFR_{last-central}}{sSFR_{last-measurement}}\right),
\end{equation}
    
\noindent and 

\begin{equation}
    \rm \delta\,C-index = log_{10}\left(\frac{\left(r_\mathrm{50,SFR}/r_\mathrm{50,rband}\right)_{\rm last-central}}{\left(r_\mathrm{50,SFR}/r_\mathrm{50,rband}\right)_{\rm last-measurement}}\right),
\end{equation}

\noindent respectively. $\delta C$-index is greater when galaxies become more concentrated. Fig.~\ref{fig:dssfrvsdcindex} shows the $\delta$sSFR as a function of $\delta C$-index for satellites at $z=0-1$. A clear correlation between $\delta$sSFR and $\delta C$-index emerges at $z\lesssim 0.7$ for HMGs, and $z\lesssim 0.5$ for IMGs. For LMGs a weak correlation is only seen at $z=0$. By $z=1$ there is no correlation between $\delta$sSFR and $\delta C$-index. We inspected this correlation up $z=2$ (not shown here) and found similar behaviours to that seen for satellites at $z=1$. {In principle, at high redshift, the small $\delta C$-index can also be interpreted as galaxy star-formation activity being fairly recent, making it hard to trace the past and current star formation. However, we show below that the sSFR radial profiles show this is not the case, and that quenching is truly acting differently at high redshift compared to low redshift. }
The results above indicate that $C$-index can be used as a proxy for ``outside-in'' quenching only in some environments and cosmic times. The emergence of a correlation between $\delta$sSFR and $\delta C$-index appears to be related to the existence of a sizeable population of satellite galaxies that have been satellites for $\gtrsim 4$~Gyr (red points in Fig.~\ref{fig:dssfrvsdcindex}). {Apart from the above trends in IMGs and HMGs, there are some galaxies in LMGs at $z<0.5$ having both decreased sSFR and increased $C$-index. Those galaxies fell in the current host a long time ago (> 4 Gyr) and tend to have smaller $R_{\rm closest-approach}$ compared to satellites in those same LMGs that do not display a correlation between $\delta$sSFR and $\delta C$-index. We think these features are more indicative of tidal interactions with the central galaxy of the groups. In fact, \citet[]{Marasco2016} analysed the H\textsc{i} morphology of galaxies in {\sc Eagle} to understand which environmental processes could be affecting them and concluded that tidal stripping becomes increasingly important relative to ram-pressure as the halo mass decreases and as the galaxies are closer to the centre of the halo. Our results appear to be consistent with this picture.} The question becomes whether satellite galaxies quench in an outside-in fashion only towards low redshift, and/or whether a combination of outside-in and inside-out quenching is taking place in high redshift satellites so the use of $C$-index as a proxy for quenching becomes limited.

To explore this question we investigate the sSFR of satellite galaxies at different cosmic epochs and environments in Fig.~\ref{fig:ssfr_profile_all}. We show separately satellites that are above and below the main sequence (for the latter we adopt two bins for mildly and well below the main sequence). Satellite galaxies above the main sequence display flat sSFR profiles in all environments and redshifts studied. For galaxies that have $\Delta\,\rm MS<-0.5$ we find that they exhibit steep profiles at $z\lesssim 0.75$ in HMGs and $\lesssim 0.3$ in IMGs. Satellites with $-0.5\le \Delta\,\rm MS<0$ appear to show clearly steep sSFR profiles at lower redshifts than galaxies with $\Delta\,\rm MS<-0.5$. Interestingly, at $z=1$ all the satellite samples have sSFR profiles consistent with being flat regardless of their $\Delta\,\rm MS$. {In addition, at $z \sim 0.7$, those below the main sequence appear to have a dip in the centre, at $r\lesssim r_{\rm 50}$. To explore this in detail, we first study the sSFR radial profiles without normalising $r$ by $r_{50}$ and find no drop in sSFR near the centre. 
By then exploring the $r_{50}$ distribution of galaxies at those high redshifts, 
we find that galaxies with the larger $r_{50}$ are those that primarily contribute to the small $r/r_{50}$ bins in Fig.~\ref{fig:ssfr_profile_all}. This brings many of the lower sSFR values of those large $r_{50}$ galaxies to the small $r/r_{50}$ bins, leading to the dip in the sSFR profiles when plotted against $r/r_{50}$. Hence, the sSFR dip cannot be interpreted as inside out quenching dominating these galaxies. This shows that sSFR profiles of galaxy populations need to be carefully analysed to be correctly interpreted.}

Previous studies have suggested that $C$-index may be less correlated with quenching at $z\approx 1$ due to quenching affecting the whole galaxy and timescales becoming very short \citep{Matharu2021}. {\sc Eagle} shows that indeed, satellite galaxies that are below the main sequence at $z>1$ have sSFR profiles that are consistent with being flat, where the SF activity of the whole galaxy decreases. At lower redshifts, however, a clear ``outside-in'' quenching signature emerges, with the sSFR in the outskirts of satellites becoming lower before the inner parts are affected. 

In the next section we discuss how quenching timescales evolve for satellite galaxies in different environments.

\begin{figure*}
	\includegraphics[width=14cm]{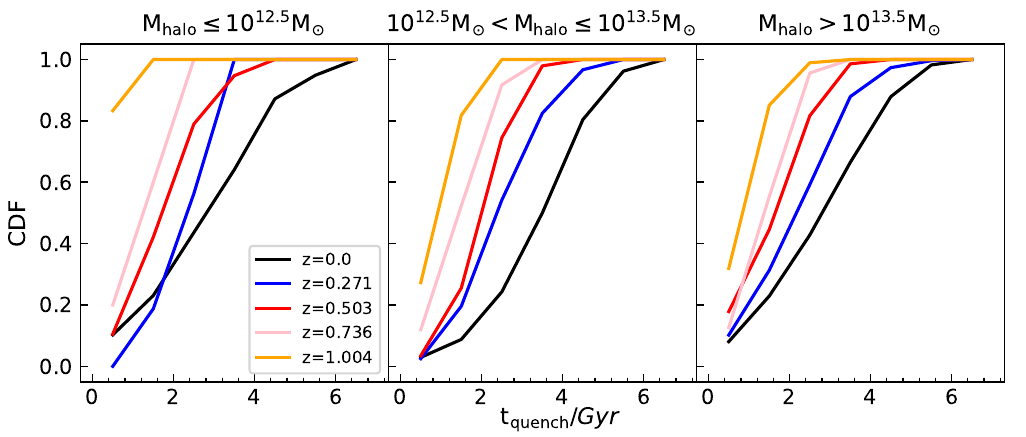}
    \caption{Cumulative distribution of $t_\mathrm{quench}$ for quenched satellites at $z\approx 0-1$ and in three host halo mass bins, as labelled. Note, $t_\mathrm{quench}$ is less than the Hubble time at each redshift, which is 13.82 Gyr, 10.58 Gyr, 8.60 Gyr, 7.13 Gyr and 5.86 Gyr from $z=0$ to $z=1$, respectively. Satellite galaxies quench on much shorter timescales with increasing redshift.}
    \label{fig:t_que_hist}
\end{figure*}

\begin{figure*}
	\includegraphics[width=14cm]{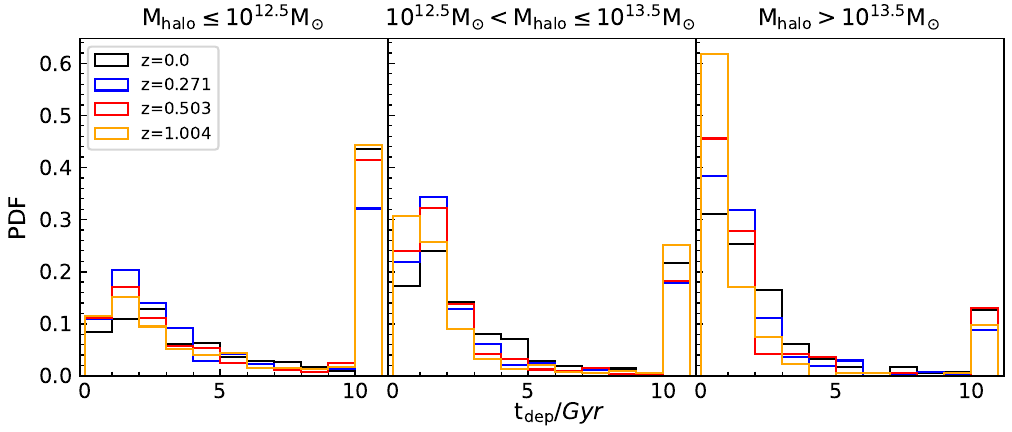}
    \caption{Probability distribution function (PDF)   of $t_\mathrm{dep}$ at redshift from 0 to 1 in three halo mass intervals, as labelled. Galaxies with $t_\mathrm{dep}$ $\ge$ 10 Gyr are considered to be in equilibrium; $\rm 1.5\,Gyr<t_\mathrm{dep}<10\,Gyr$ galaxies are slowly depleting their ISM; and $t_\mathrm{dep}\le 1.5$~ Gyr are rapidly depleting their ISM. There is a greater fraction of equilibrium galaxies in LMGs than in IMGs and HMGs, and $t_\mathrm{dep}$ on average increases with decreasing redshift. }
    \label{fig:t_dep_hist}
\end{figure*}

\begin{figure}
	\includegraphics[width=\columnwidth]{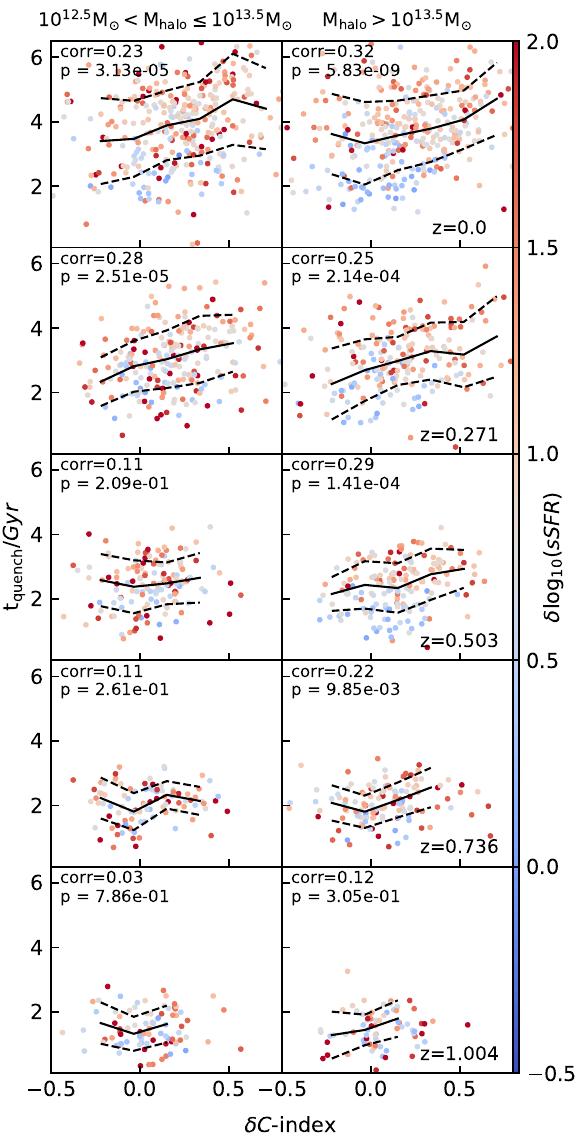}
    \caption{The $t_{\rm quench}$ versus $\delta C$-index for {\sc Eagle} galaxies colour-coded by $\delta$sSFR at redshifts $0-1$, as labelled. $\delta$sSFR and $\delta\,C$-index is the same as Fig.\ref{fig:dssfrvsdcindex}. The medians and 1~$\sigma$ percentile ranges are as solid and dashed lines, respectively. In each panel, we also show the Spearman correlation coefficient and $p$-value. The LMG interval is not plotted because of the limited number of quenched galaxies as in Table.~\ref{tab:tquenchtable}.  
    At $z\le 0.5$ satellite galaxies in IMGs and HMGs exhibit a clear correlation between $t_{\rm quench}$ versus $\delta C$-index (longer $t_{\rm quench}$ is correlated with more concentrated star formation) that is not present at $z=1$. }
    \label{fig:dque_cindex}
\end{figure}

\begin{table}
	\centering
	\caption{{The median $t_{\rm quench}$ (in units of Gyr) with standard errors and the number of galaxies for {\sc Eagle} satellite galaxies in each halo mass interval at different redshifts. ``Num'' refers to the number of galaxies in each sample.}}
	\label{tab:tquenchtable}
	\begin{tabular}{ccccc} 
		\hline
		$z$ & LMG & IMG & HMG & All\\
		$ $  & ${\le} 10^{12.5}M_{\odot}$ & $ 10^{12.5-13.5}M_{\odot}$ &${>} 10^{13.5}M_{\odot}$ & $ $ \\
		\hline
		$z$=$0$ & 3.11$\pm$0.26 & 4.00$\pm$0.07 & 3.32$\pm$0.07 & 3.66$\pm$0.05 \\
            Num    & 40 & 343 & 399 &  782\\
            \hline
		  $z$=0.271 & 3.06$\pm$0.20 & 2.98$\pm$0.07 & 2.72$\pm$0.07 & 2.92$\pm$0.05  \\
            Num    & 16 & 240 & 265 & 521\\
            \hline
		$z$=0.503 & 2.09$\pm$0.22 & 2.38$\pm$0.07 & 2.24$\pm$0.06 & 2.36$\pm$0.05 \\
            Num    & 19 & 145 & 213 & 377\\
            \hline
            $z$=0.736 & 1.75$\pm$0.17 & 1.96$\pm$0.07 & 1.95$\pm$0.05 & 1.93$\pm$0.05 \\
            Num    & 10 & 109 & 159 & 278 \\
            \hline
            $z$=1.004 & 0.97$\pm$0.15 & 1.50$\pm$0.06 & 1.21$\pm$0.07 & 1.22$\pm$0.05 \\
            Num    & 6 & 99 & 94 & 199 \\
            \hline
            $z$=2.012 & 0.41$\pm$0.10 & 1.00$\pm$0.08 & - & 0.92$\pm$0.08 \\
            Num    & 3 & 16 & - & 19\\
		\hline
	\end{tabular}
\end{table}

\begin{table}
	\centering
	\caption{{As Table~\ref{tab:tquenchtable} but for $t_{\rm dep}$.}}
	\label{tab:tdeptable}
	\begin{tabular}{ccccc} 
		\hline
		$z$ & LMG & IMG & HMG & All\\
		$ $  & ${\le} 10^{12.5}M_{\odot}$ & $ 10^{12.5-13.5}M_{\odot}$ &${>} 10^{13.5}M_{\odot}$ & $ $ \\
		\hline
		$z$=$0$ & 6.60$\pm$0.16 & 2.53$\pm$0.16 & 1.66$\pm$0.17 & 2.93$\pm$0.09 \\
            Num    & 649 & 922 & 371 &  1942\\
            \hline
		  $z$=0.271 & 3.51$\pm$0.21 & 1.76$\pm$0.12 & 1.21$\pm$0.17 & 1.87$\pm$0.09  \\
            Num    & 358 & 868 & 305 & 1531\\
            \hline
		$z$=0.503 & 4.86$\pm$0.25 & 1.73$\pm$0.25 & 1.06$\pm$0.26 & 1.91$\pm$0.12 \\
            Num    & 280 & 661 & 169 & 1110\\
            \hline
            $z$=1.004 & 6.48$\pm$0.20 & 1.63$\pm$0.20 & 0.79$\pm$0.15 & 2.02$\pm$0.11 \\
            Num    & 454 & 796 & 175 & 1425 \\
            \hline
	\end{tabular}
\end{table}

\subsection{Quenching timescales and their relation to $C$-index}

To quantify quenching timescales, we take $t_\mathrm{quench}$, which is the time between the last time satellite galaxies resided in the sSFR MS and the first time they became quenched. 
Note we select galaxies so that there is no overlap between the quenched sample at different redshifts; if a galaxy quenched at $1<z<2$ it will only be included in the sample at $z=1$. {The number of galaxies in each sample and their median $t_\mathrm{quench}$ with standard errors are listed in Table.~\ref{tab:tquenchtable}. }{Additionally, we also use $t_\mathrm{dep}$, which is a timescale that quantifies how long it would take to exhaust the ISM of galaxies given present rates of inflow, outflow and star formation. Median $t_\mathrm{dep}$ values with standard errors are listed in Table.~\ref{tab:tdeptable}. } These quantities are defined in $\S$~\ref{Esec:time_metric}.

{The use of $t_\mathrm{quench}$ and $t_\mathrm{dep}$ have advantages and limitations. The number of quenched galaxies for which we measure $t_{\rm quench}$, could be a small fraction of the total number of satellites in some samples; e.g. $\sim$ 4 per cent galaxies in LMGs at $z=0$ $t_\mathrm{quench}$ values could represent only a small sample of quenched galaxies. In these cases $t_\mathrm{quench}$ would offer a biased view of the satellite population. On the contrary, $t_\mathrm{dep}$ can only be calculated for galaxies that have gas ISM and SFR > 0 . Even though $t_\mathrm{dep}$ can be obtained for most galaxies ($>50$\%), it can change significantly if the rates of inflow/outflow/star formation change - something that is expected for satellite galaxies \citet{Wright2022}. Therefore, we opt to study both $t_\mathrm{quench}$ and $t_\mathrm{dep}$ which should give us a more complete picture of satellite galaxies in the simulation.}

Firstly, we investigate the cumulative distribution of $t_\mathrm{quench}$ at $z=0-1$ in three halo mass intervals in Fig.~\ref{fig:t_que_hist}. Both { Fig.~\ref{fig:t_que_hist} and Table~\ref{tab:tquenchtable} show that satellites in LMGs have similar or slightly lower $t_\mathrm{quench}$ than those in IMGs and HMGs. However, the number of galaxies contributing to $t_\mathrm{quench}$ for LMGs is small, as most satellites in these groups do not quench. 

The picture of LMGs changes when we study $t_\mathrm{dep}$ instead as seen in Fig.~\ref{fig:t_dep_hist} and Table~\ref{tab:tdeptable}. A large fraction of galaxies in LMGs are in an equilibrium mode ($t_\mathrm{dep}$ $\ge$ 10 Gyr), where gas inflow, outflow and star formation rates are stable (which is the case for most field galaxies too) and only a small fraction of galaxies are in a rapidly depleting ISM phase ($t_\mathrm{dep}$ $\le$ 1.5 Gyr). This confirms $t_\mathrm{quench}$ is low in LMGs relative to IMGs and HMGs, because it is simply reflecting the most extreme satellites that do quench in these halos. The large fraction of equilibrium galaxies explains why in LMGs both $\delta$sSFR and $\delta C$-index are smaller than the values found in IMGs and HMGs, regardless of redshift.

Fig.~\ref{fig:t_dep_hist} shows a trend of $t_{\rm dep}$ decreasing from LMGs to IMGs to HMGs. The emerging picture is that in HMGs, satellites that are quenched did that in shorter timescales than galaxies in other environments, and those that are in the process of quenching are doing that faster than satellites in IMGs/LMGs, with shorter ISM depletion times.} This is not surprising as the efficiency of RPS as an environmental process affecting a galaxy's gas reservoir also increases with halo mass in {\sc Eagle} \citep{Marasco2016}.

{Furthermore, we see $t_\mathrm{quench}$ and $t_\mathrm{dep}$ evolve with redshift. }In all environments, there is the same qualitative trend of $t_\mathrm{quench}$ becoming on average longer from $z=1$ to $z=0$. Satellite galaxies have a wide range of $t_\mathrm{quench}$ from 1 to 6.5 Gyr at $z=0$, while at $z=1$ they all have $\lesssim 2.5$~Gyr (with a median of $1.22 \pm 0.05$~Gyr). At $z=2$, $t_{\rm quench}$ is much shorter with a median of $0.92 \pm 0.08$~Gyr. 
A Kolmogorov$-$Smirnov (K$-$S) test is applied on the $t_\mathrm{quench}$ distributions in all environments at different redshifts compared to $z = 0$. In all cases, we find $p$-values$\ll 10^{-20}$, which allows us to reject the null hypothesis the two distributions are the same.

{Similarly, the fraction of rapidly depleting galaxies increases with increasing redshift. At $z$=1, over $50$\% of galaxies are in the regime of rapid ISM depletion in HMGs (in the $t_\mathrm{dep}$ 0-1 Gyr bin). This happens as these galaxies are undergoing significant gas stripping. This had already been discussed in \citet{Wright2022}, but here we find that these rapid depletion timescales are associated with flat sSFR radial profiles consistent with SF activity decreasing across the whole galaxy. This gas depletion timescale becomes longer towards $z=0$ with a median of $\approx 2$~Gyr for HMGs. These trends are qualitatively consistent with the ones reported here for $t_{\rm quench}$. 
}

Because the universe is younger at $z=1$ it is not surprising that quenching timescales are shorter than at $z=0$. However, when we compare the distributions with the age of the universe ($t_{\rm age}$) at each redshift for the cosmology adopted by {\sc Eagle}, we find that the $z=1$ quenching timescales are shorter ($\approx 0.20\,t_{\rm age}$) than those at $z=0$ ($\approx 0.27\,\rm t_{\rm age}$), on average, showing that relative to the universe's age, one can conclude that satellite galaxies quench faster at high redshift than in the local universe in {\sc Eagle}. This is likely related to the efficiency of RPS being higher at $z=1$ than at $z=0$ in high density environments, as shown by \citet{Bahe2014}.

We hypothesize that short quenching timescales at high redshift lead to changes in the sSFR radial profiles that affect the whole galaxy rather than showing an ``outside-in'' quenching signature. This would lead to important changes in the galaxy sSFR (i.e. large $\delta$sSFR), but little in $C$-index. The lengthier quenching timescales towards lower redshifts are associated with changes in sSFR radial profiles that are characterised by a larger decrease in the outskirts' sSFR than in the inner parts of galaxies (a.k.a. ``outside-in'' quenching). We test our hypothesis by using the values of $\delta$sSFR and $\delta C$-index introduced in $\S$~\ref{Esec:ssfrtimeprofile}, which can be computed for all satellite galaxies (quenched or star-forming) and comparing these with $t_{\rm quench}$ for quenched satellites. Because the sample of quenched satellites at $z=2$ is so small, we do not show them in the figure. This is shown in Fig.~\ref{fig:dque_cindex} separately for quenched satellites in different environments and cosmic times.
We do not show LMGs here as there are only a small number of quenched satellites in these environments, insufficient to draw any conclusion. 

Apart from the existing correlation between $\delta$sSFR and $\delta C$-index that manifests at low redshift, there is a positive relation between $t_\mathrm{quench}$ and $\delta C$-index that is clear at $z \lesssim 0.5$. This correlation becomes stronger with time as seen by the Spearman correlation coefficient becoming larger towards $z=0$. Another interesting trend is that the correlation between $t_{\rm quench}$ and $\delta C$-index is always stronger (see Spearman coefficients)  for satellites in HMGs than in IMGs at fixed redshift. In fact, a correlation between $t_{\rm quench}$ and $\delta C$-index starts to emerge only at $z\lesssim 0.27$ for IMGs, while for HMGs, this happens at $z<0.7$.

\section{Discussion}\label{Esec:discussion}

We aim to better understand the SF quenching driven by environment using the {\sc Eagle}/C-{\sc Eagle} hydrodynamical simulations.
By comparing the concentration index [$C$-index, log$_{10}$($r_\mathrm{50,SFR}/r_\mathrm{50,rband}$)] from the simulations with observations from the SAMI survey at $z=0$, we test how the extent of SF in galaxies depends on environment (Low-Mass Groups, LMGs, $M_{200}$ $\leq 10^{12.5} M_{\odot}$; Intermediate-Mass Groups, IMGs, $10^{12.5} M_{\odot}< M_{200} \leq 10^{13.5} M_{\odot}$; High-Mass Groups, HMGs, $M_{200}$ > $10^{13.5} M_{\odot}$; and clusters, $M_{200}$ > $10^{14} M_{\odot}$), and confirm that the $C$-index is a parameter well-suited to identify galaxies that are undergoing ``outside-in'' quenching. Furthermore, we use {\sc Eagle} to test the relationship between SFR and $C$-index across cosmic time to understand how environmental quenching takes place. {\sc Eagle} shows the way satellite galaxies quench changes with cosmic time in a way that clear ``outside-in'' quenching signatures emerge only towards low redshift (with the exact time depending on environment). Thus, $C$-index as a quenching proxy becomes less useful with increasing redshift. Due to this time evolution, below we first discuss the quenching of satellite galaxies at $z=0$ in $\S$~\ref{Esec:discussion:z0}, followed by a discussion of its time evolution ($\S$~\ref{Esec:discussion:highz}).

\subsection{Quenching of satellites in the local Universe}\label{Esec:discussion:z0}

Both SAMI and {\sc Eagle}/C-{\sc Eagle} show a higher fraction of SF-concentrated galaxies in higher mass halos. Along with the fact that the SF-concentrated galaxies sit beneath the SFR-$M_{\star}$ MS, lower star formation in the discs is more likely to be associated with a low $C$-index, which could be seen as a smoking gun of ``outside-in'' quenching  \citep[e.g.][]{Koopmann_2006}. HMGs and galaxy clusters are environments where RPS will efficiently expel the gas from the disc, especially in the outskirts where the gas is less bound, resulting in low or no star formation in the outer regions of galaxies. \citet[]{Cortese2012} found the star formation in the inner parts of satellite galaxies in the Galaxy Evolution Explorer \citep{Martin2005} GR6 data and Virgo cluster is significantly less affected by the removal of gas due to RPS. \citet[]{Wang2022} using SAMI argue that the $C$-index is a measurement that can be used to select galaxies currently undergoing environmental quenching in an outside-in fashion, which we confirm here using the {\sc Eagle}/C-{\sc Eagle} simulations. At low redshift, there are other observational works showing evidence of environment suppressing star formation in satellite galaxies in an outside-in fashion in groups and clusters \citep[e.g.][]{Bretherton2013, Schaefer2017}. 

For {\sc Eagle} we use orbital parameters and the time a galaxy has been a satellite of its current host, $t_\mathrm{sat}$ ($\S$~\ref{Esec:time_metric}), to understand environmental quenching. Galaxies that recently fell into the group/cluster or have orbits with large pericentric radii (i.e. $\gtrsim 0.7\,r_{200}$) are less affected by their environment (at least as seen by $C$-index). In denser regions, there is a stronger negative relationship between the $C$-index and $t_\mathrm{sat}$. It shows the longer a galaxy is a satellite, more gas can be stripped off by the ICM. Also, while the present  location (and by extension the current phase-space location), contains some information on the satellite infall timescale, there is also a large scatter in the orbital histories of galaxies at a given radius. The evidence of this is the fact that $C$-index is less correlated with the current distance to the halo centre, $R_{\rm current}$, but more correlated with $N$-pericentre and $R_{\rm closest-approach}$ ($\lesssim 0.8 r_{200}$) - see Fig.~\ref{fig:CvsMnperi}. {\citet[]{Wang2022} studied the phase-space diagram of SAMI satellite galaxies to investigate the relationship between their current location and C-index. They did not find a significant relationship between the current location and C-index, consistent with the simulation results.} \citet[]{Arthur2019, DiCintio2021} and \citet{Coenda2021} reported a dependence of quenching on orbital parameters which agrees with our simulation findings. 

We track the sSFR and $C$-index histories of satellite galaxies at $z=0$ (Fig.~\ref{fig:sfr_cindex_vst_s}), and find their sSFR drops quicker as we  move from LMGs, IMGs to HMGs. We also find $C$-index evolves similarly to sSFR; when the sSFR has not dropped off the MS, the $C$-index changes little, while large drops in sSFR are accompanied by a large decrease in $C$-index. Another piece of supporting evidence is that when the satellite galaxy sample is split by $\Delta$MS, satellites with $\Delta\,\rm MS\le -0.5$ show a clearly steeper sSFR radial profile than galaxies with $\Delta\,\rm MS>0$.

The $\Delta$MS, $C$-index time evolution and $\delta$sSFR versus $\delta C$-index diagram (Fig.~\ref{fig:dssfrvsdcindex}) show longer $\delta t_\mathrm{last-central}$ (>5Gyr) lead to lower $C$-index and decreasing sSFR at $z \sim$ 0. 
We find $t_\mathrm{quench}$ timescales at $z$ = 0 display a wide range with values even as high as $6.5$~Gyr, with a median of $3.66 \pm 0.05$~Gyr. 
{From our analysis of $t_\mathrm{dep}$, we find equilibrium satellite galaxies are primarily found in LMGs. The fraction of galaxies with $t_\mathrm{dep}$ < 1 Gyr gradually increases with increasing host halo mass, which is quantitatively consistent with $t_\mathrm{quench}$ timescales. }
The results we obtain from {\sc Eagle} agree with the study of \citet[]{Finn2018}, who built a model to constrain the star-forming disk shrinking timescale in the cluster environment, and found galaxies in clusters that undergo ``outside-in'' RPS are more likely to have long quenching timescales ($\gtrsim 2$~Gyr).

Many studies have shown that galaxies may have started or even finished their star-formation quenching before they fall into the current host halo, a process often referred to as pre-processing \citep[e.g.][]{Mihos2004, Oh_2018}. To investigate the effect of pre-processing, we select satellites that have $t_\mathrm{last-central}$ > $t_\mathrm{sat}$ and previous host halo mass, $M_{200}$ $ > 10^{12.5}\,\rm M_{\odot}$) following the selection in \citet[]{Wright2022}. With this selection, there are 3, 229, 137 galaxies in LMGs, IMGs, and HMGs, respectively, that have been pre-processed. If we track these galaxies in snapshots prior to $t_\mathrm{sat}$, we find that they show some decrease in sSFR but no significant change in $C$-index. This is broadly consistent with what we see at high redshift (discussed in the next section). Other studies also show pre-processing has only a mild effect in group environments \citep[e.g.][]{Vijayaraghavan2013, Hou2014} in agreement with what we see in {\sc Eagle}. Therefore, throughout our study, we do not exclude pre-processed galaxies as they do not bias the overall results discussed above. 

\subsection{Quenching of satellites across cosmic time}\label{Esec:discussion:highz}

At intermediate redshifts ($z=0.27-0.5$), we see similar trends to those found at $z=0$, namely, galaxies in dense environments tend to have more concentrated SF. SF-concentrated galaxies are still located beneath the MS of SFR-$M_{\star}$ diagram, which supports the idea that environmental quenching acts in an ``outside-in'' fashion at intermediate redshifts. Similar results in observations carried out with the Very Large Telescope at intermediate redshifts have also been found in \citet[]{Bamford2007} and \citet{Boesch2013}. \citet[]{Jaffe2011} used cluster galaxies at $0.3<z<0.9$ and found that gas discs in cluster galaxies have been truncated which leads to star formation being more concentrated in clusters than in the field. This agrees with the findings in \citet[]{Vaughan2020} of cluster galaxies at $z\sim 0.5$ having smaller H$\alpha$ to stellar continuum size ratios than field galaxies, again supporting the ``outside-in'' quenching scenario. {Although there are studies showing environmental quenching to be overly efficient at quenching low-mass satellites (stellar masses $\approx 10^9\,\rm M_{\odot}$) in simulations of resolution similar to {\sc Eagle} \citep[e.g.][]{Bahe2017, Kukstas2022},  the general ``outside-in'' quenching scenario we obtain in {\sc Eagle} and C-{\sc Eagle} is consistent with SAMI, and clearly present in satellite galaxies of stellar masses $\gg 10^9\,\rm M_{\odot}$. These results give us confidence that the simulation suite used here is good enough to offer important insights into environmental quenching. }

In {\sc Eagle} we find that $t_{\rm quench}$ is correlated with $\delta C$-index at intermediate redshifts, but the correlation is weaker than at $z=0$. Satellite galaxies at 0$\lesssim z \lesssim$ 0.5 take longer to quench ($t_{\rm quench}$ has a median value of 3.66 $\pm$ 0.05 Gyr, 2.92 $\pm$ 0.05 Gyr and 2.36 $\pm$ 0.05 Gyr at $z$=0, 0.271 and 0.503, respectively) than satellites at 0.75 $\lesssim z \lesssim$ 1 ($t_{\rm quench}$ has a median value of 1.93 $\pm$ 0.04 $\pm$ 0.04 Gyr and 1.22 $\pm$ 0.05 Gyr at $z=0.736$ and $z=1.004$). {Additionally, there are more galaxies that show rapid ISM depletion in IMGs and HMGs with increasing redshifts, similar results have been found in \citet[]{NogueiraCavalcante2018}. \citet[]{Wright2022} found $t_\mathrm{dep}$ is highly related to the orbits of satellites; galaxies at each peri-centre will have an increasing outflow rate leading to shorter $t_\mathrm{dep}$. 
There is less of a trend of $t_{\rm dep}$ with redshift  in LMGs due to environmental effects in low density environments being weak.
}

Both the $C$-index relation with the position of galaxies in the SFR-$M_{\star}$ diagram (Fig.~\ref{fig:sfrvsm}), and  $\delta$sSFR versus $\delta C$-index diagram (Fig.~\ref{fig:dssfrvsdcindex}) mostly disappear at 0.75 $\lesssim z \lesssim$ 2 (we explicitly showed this for up to $z=1$ but corroborated that this was the case up to $z=2$ - too few satellite galaxies exist at $z>2$ in {\sc Eagle} to carry out a statistical study of them). Towards $z=1-2$, satellite galaxies have a much narrower $C$-index range, and fewer SF-concentrated galaxies tend to be found beneath the MS of the SFR-$M_{\star}$ diagram (Fig.~\ref{fig:sfrvsm}). At $z=1-2$, the positive relation between $\delta$sSFR and $\delta C$-index is almost absent (Fig.~\ref{fig:dssfrvsdcindex}). 

In addition, at $z=1-2$ in {\sc Eagle}, we do not see an increasing fraction of SF-concentrated galaxies in HMGs compared to LMGs and IMGs. Galaxies in all environments seem to have extended star formation. 
Furthermore, we find that satellites whose sSFR has significantly decreased in their time as satellite galaxies still have high $C$-index, so whatever star formation they have is still extended (Fig.~\ref{fig:sfrvsm}, \ref{fig:dque_cindex}). These galaxies show sSFR profiles consistent with having SF suppression across the whole galaxy and short gas depletion and quenching timescales. This implies that the fast quenching these high-$z$ satellites experience is so violent that no signature of ``outside-in'' quenching is left and $C$-index is unchanged. 
Our results are consistent with the observations of \citet[]{Matharu2021} who used a similar $r_\mathrm{50,H\alpha}/r_\mathrm{50,cont}$ ratio to characterise the extent of SF in cluster galaxies at $z \sim$ 1 and found cluster galaxies to have consistent $C$-index with field galaxies. When they analysed post-starburst cluster galaxies they found that their stellar populations had signatures of having been quenched in an outside-in fashion. They concluded that quenching must be rapid in cluster galaxies to not see changes in $C$-index in current SF satellites but ``outside-in'' quenching signatures in recently quenched satellites.

Many studies have inferred from observations that the star-formation quenching at $z \sim$ 1 is rapid, with shorter quenching timescales compared to lower redshift galaxies \citep[e.g.][]{Newman2014, Foltz_2018, vanderBurg2020}. Using deep Gemini Multi-Object Spectrograph-South spectroscopy, \citet[]{Mok2013} found group galaxies at $z \sim$ 1 have a delay between being accreted in their current host and the onset of truncation in star formation of $< 2$~Gyr, shorter than the $3-7$~Gyr inferred at $z = 0$ \citep{Wetzel2012}. They also found the actual quenching process also occurs quickly, with an exponential decay timescale $< 1$~Gyr. \citet[]{Balogh2016} used galaxies in groups and clusters at $0.8<z<1.2$ and found that quenching timescales appeared to be much shorter than at $z$=0, with expulsion via modest outflows and strangulation suggested as likely processes dominating the quenching of these $z\approx 1$ group galaxies. \citet[]{Papovich2018} found starvation combined with rapid gas depletion and ejection at $z\gtrsim 1$ to be the dominant form of environmental quenching. They suggest these processes become less efficient with time, in a way that at $z<0.5$ RPS becomes the more dominant process behind the quenching of satellites in clusters. Pre-processing has also been found to be acting in high redshift galaxies \citep[e.g.][]{Nantais2016, Werner2021}. \citet[]{Foltz_2018} suggested that the delay time of environmental quenching decreases with both increasing redshift and host halo mass in the `delayed-then-rapid' quenching scenario. \citet[]{Wright2022} used {\sc Eagle} to analyse gas flows in satellite galaxies at high redshift, and found that there is a greater fraction of galaxies undergoing rapid gas depletion compared to satellites at $z=0$, with the expectation being that their ISM reservoir would exhaust in $<1$~Gyr.

In our study we find that the sSFR radial profiles of satellite galaxies at $z=1$ are consistent with being flat, even when these galaxies have $\Delta\,\rm MS<-0.5$ (Fig.~\ref{fig:ssfr_rprofile}). This is consistent with processes other than RPS playing an important role in quenching satellite galaxies at high redshift. 
Therefore, we can conclude that galaxies at $z$=1 experience rapid environmental quenching, with the actual quenching timescale being short compared to $z$=0.

\section{Conclusions} \label{Esec:conclusion}

By comparing the {\sc Eagle}/C-{\sc Eagle} simulations at $z$=0 with the SAMI Galaxy Survey, and exploring the sSFR radial profiles of satellite galaxies in the simulations, we find that the star formation concentration index ($C$-index) is a good proxy to select galaxies that are currently undergoing environmental star-formation quenching. We use the $C$-index to study how star-formation quenching depends on the environment (including galaxy groups and clusters). By applying the same $C$-index at a higher redshift to simulations, we explore how star-formation quenching evolves towards $z$=2. Along with the $C$-index, we explore several timescale parameters and orbital information ($\S$~\ref{Esec:orbit}) in the {\sc Eagle} simulation to understand how galaxies evolve before and after they fall into the current host, and how that evolution is traced by changes in $C$-index. 

In this study, we analyse satellite galaxies in four halo mass bins: LMGs ($M_{200}$ $\leq 10^{12.5} M_{\odot}$), IMGs ($10^{12.5} M_{\odot}< M_{200} \leq 10^{13.5} M_{\odot}$), HMGs ($M_{200}$ > $10^{13.5} M_{\odot}$) and clusters ($M_{200}$ > $10^{14} M_{\odot}$). Our sample is separated into two types of galaxies: regular galaxies with $C$-index $\ge -0.2$ and SF-concentrated galaxies with $C$-index $< -0.2$. Our conclusions are summarised below:

\begin{enumerate}

\item At $z = 0$, in the {\sc Eagle} and C-{\sc Eagle} simulations, there is a larger fraction of SF-concentrated galaxies in HMGs/clusters compared to IMGs and LMGs, with the fraction of SF-concentrated galaxies being 22$\pm$2\%, 27$\pm$1\%, 39$\pm$2\% and 51$\pm$2\% for LMGs, IMGs, HMGs and clusters, respectively. The simulation results agree qualitatively with the SAMI results within, except for the overall smaller fraction of SF-concentrated galaxies in SAMI (10$\pm$3\%, 13$\pm$4\%, 28.8$\pm$4\% and 29.5$\pm$4\% for LMGs, IMGs, HMGs and galaxies in HMGs within the cluster halo mass interval, respectively). This tension remains even if we match the stellar mass distribution between SAMI and {\sc Eagle}/C-{\sc Eagle}. Regardless, the tension is small, and the trends seen in the simulations are broadly consistent with those of SAMI.

\item At $z = 0$, galaxies with a larger number of pericentric passages ($N$-pericentre) tend to have lower $C$-indices in IMGs and HMGs, and even at fixed sSFR, galaxies with a higher $N$-pericentre tend to have a smaller $C$-index. In HMGs, $C$-index is better correlated with t$_\mathrm{sat}$ and $R_{\rm closest-approach}$ than the current radius $R_{\rm current}$. The difference with $N$-pericentre is consistent with environmental processes such as RPS being more efficient towards the centres of clusters where the ICM density is higher. {\sc Eagle} predicts the $C$-index to be a better proxy for how long satellites have been in the current host than their current position within the cluster, at least for SF satellite galaxies.

\item The sSFR radial profiles (Fig.~\ref{fig:ssfr_rprofile}) show that SF-concentrated galaxies have very little SF in their outskirts in HMGs and clusters, which is consistent with the ``outside-in'' quenching scenario. In HMGs, both the sSFR and $C$-index of satellite galaxies steeply decline with time. These satellites also take a shorter timescale ($\sim$ 3Gyr) to leave the MS compared to satellites in LMGs/IMGs ($\sim$5Gyr), indicating that the $C$-index is a powerful property to study the physical processes behind quenching in $z=0$ satellite galaxies.

\item The range of $C$-index values in satellite galaxies narrows down with increasing redshift in {\sc Eagle}. Changes in $C$-index ($\delta C$-index) and sSFR ($\delta$sSFR) are correlated in HMGs only at $z\lesssim 0.7$ and  $z\lesssim 0.5$ for IMGs. $C$-index shows very little change as galaxies become satellites at $z\gtrsim 0.7$ to $z=2$. In fact, we see that these high-z satellites have sSFR radial profiles consistent with quenching suppressing SF across the whole galaxy, rather than preferentially in the outskirts (as $z<0.5$ satellites show).

\item By studying quenched and quenching satellite galaxies in {\sc Eagle} at different redshifts and environments,
{we see that quenching timescales decrease from LMGs, IMGs, to HMGs. This suggests environmental quenching
caused is quicker in denser environments, as expected.} we find that $t_\mathrm{quench}$ decreases significantly with increasing redshift. At $z$= 0, 0.271, 0.503, 0.736, 1.004 and 2, $t_\mathrm{quench}$ has a median value of 3.66 $\pm$ 0.05 Gyr, 2.92 $\pm$ 0.05 Gyr, 2.36 $\pm$ 0.05 Gyr, 1.93 $\pm$ 0.04 $\pm$ 0.04 Gyr, 1.22 $\pm$ 0.05 Gyr, and $0.92 \pm 0.08$~Gyr, respectively. {The $t_\mathrm{dep}$ distribution shows a qualitatively similar trend, with the fraction of rapidly depleting galaxies increasing at high redshift.}
Fast environmental quenching at high $z$ is associated with SF suppression across the whole galaxy, which likely reflects the fact that gas loss is very fast \citep{Wright2022} and that mechanisms such as starvation and strangulation are more efficient at high $z$, with RPS dominating later on in the evolution of satellite galaxies. The latter is what gives rise to the clear ``outside-in'' quenching signature seen in satellite galaxies in {\sc Eagle} at $z\lesssim 0.5$.

\end{enumerate}

Our work shows that environmental quenching at low redshift leaves clear ``outside-in'' quenching signatures with a slow quenching timescale > 2Gyr. We also show that the way environmental quenching takes place evolves with redshift. Galaxies at high redshift have short quenching timescales, < 2Gyr, and have sSFR radial profiles consistent with being flat but offset down compared to normal SF galaxies. 

\section*{Data Availability}

The {\sc Eagle} simulations are publicly available; see \citet{McAlpine2016, EAGLE17} for how to access {\sc EAGLE} data. The C-{\sc Eagle} data are available at \url{https://ftp.strw.leidenuniv.nl/bahe/Hydrangea/}. 

All observational data presented in this paper are available from Astronomical Optics' Data Central service at \url{https://datacentral.org.au} as part of the SAMI Galaxy Survey Data Release 3 \citep{Croom_2021}.

\section*{Acknowledgements}

CL has received funding from the Australian Research Council Centre of
Excellence for All Sky Astrophysics in 3 Dimensions (ASTRO 3D), through project number CE170100013, and the Australian Research Council Discovery Project (DP210101945). JvdS acknowledges support of an Australian Research Council Discovery Early Career Research Award (project number DE200100461) funded by the Australian Government. YMB gratefully acknowledges funding from the Netherlands Organization for Scientific Research (NWO) through Veni grant number 639.041.751. DW acknowledges support of scholarship under State Scholar Fund from the China Scholarship Council (CSC). 

This work made use of the supercomputer OzSTAR which is managed through the Centre for Astrophysics and Supercomputing at Swinburne University of Technology. This supercomputing facility is supported by Astronomy Australia Limited and the Australian Commonwealth Government through the national Collaborative Research Infrastructure Strategy (NCRIS). 

This work used the DiRAC@Durham facility managed by the Institute for Computational Cosmology on behalf of the STFC DiRAC HPC Facility (\url{www.dirac.ac.uk}). The equipment was funded by BEIS capital funding via STFC capital grants ST/K00042X/1, ST/P002293/1 and ST/R002371/1, Durham University and STFC operations grant ST/S003908/1. DiRAC is part of the National e-Infrastructure.

This equipment was funded by BIS National E-infrastructure capital grant ST/K00042X/1, STFC capital grant ST/H008519/1, and STFC DiRAC Operations grant ST/K003267/1 and Durham University. DiRAC is part of the National E-Infrastructure.
We acknowledge the Virgo Consortium for making
their simulation data available. The {\sc Eagle} simulations were performed using the DiRAC-2 facility at
Durham, managed by the ICC, and the PRACE facility Curie based in France at TGCC, CEA, Bruyeres-le-Chatel. The {\sc C-EAGLE} simulations were in part performed on the German federal maximum performance computer ``HazelHen'' at the maximum performance computing centre Stuttgart (HLRS), under project GCS-HYDA / ID 44067 financed through the large-scale project ``Hydrangea'' of the Gauss Center for Supercomputing.

The SAMI Galaxy Survey is based on observations made at the Anglo-Australian Telescope. SAMI was developed jointly by the University of Sydney and the Australian Astronomical Observatory (AAO). The SAMI input catalogue is based on data taken from the Sloan Digital Sky Survey, the GAMA Survey and the VST ATLAS Survey. The SAMI Galaxy Survey is supported by the Australian Research Council (ARC) Centre of Excellence ASTRO 3D (CE170100013) and CAASTRO (CE110001020), and other participating institutions. The SAMI Galaxy Survey website is \url{http://sami-survey.org/}.



\bibliographystyle{mnras}
\bibliography{PHD_ref} 



\appendix

\section{$C$-index comparison} \label{Esec:cindex_compare}

\begin{figure}
	\includegraphics[width=\columnwidth]{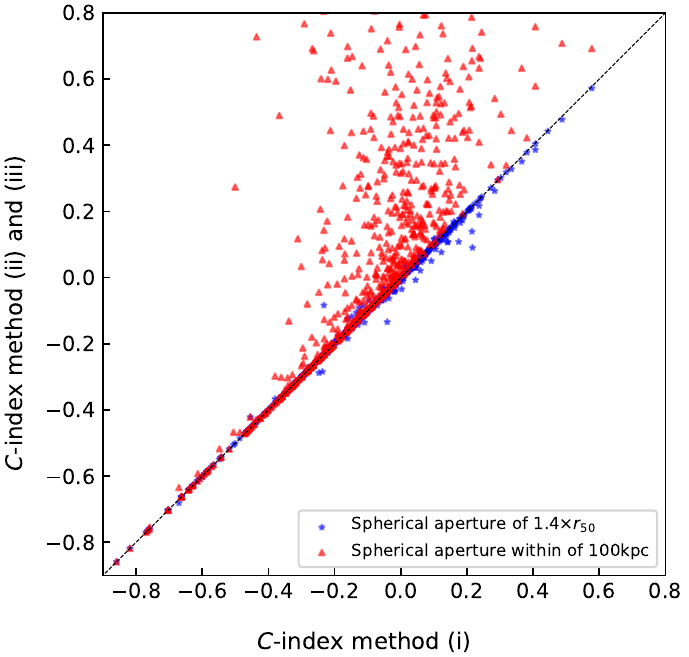}
    \caption{$C$-index comparison between 3 methods. The x-axis shows $C$-index calculated by only including gas particles within a cylinder of radius 1.4 $r_{50}$ and height 0.5 $r_{50}$ (method (i)). The y-axis shows $C$-index calculated using methods (ii) (blue) and (iii) (red), described in the text, as labelled. The one to one line is shown for reference.}
    \label{fig:cindexcompare}
\end{figure}

In {\sc Eagle} and C-{\sc Eagle}, we can directly calculate $r_{\rm 50, SFR}$ based on the star formation rate associated with each gas particle. We consider SAMI's observational limitations and size selection: for the median redshift and half-light radius of the galaxies observed, the $15''$ aperture corresponds to $\approx 1.4\,\rm r_{50}$. Therefore, we apply the 1.4 $r_{50}$ aperture on the simulated galaxies to exclude particles that would not fall within the SAMI optical IFU. 
To test how these different choices impact the measured $C$-index, we introduce 3 methods of calculating $C$-index in the simulated galaxies:

\begin{enumerate}
  \item Include all gas particles within a cylinder size of radius 1.4 $r_{50}$ and height $0.5\,\rm r_{50}$ (this is the default measurement we use in this paper).  
  \item Include all gas particles within a spherical aperture of radius 1.4 $r_{50}$.
  \item Include all gas particles within a spherical aperture of radius 100~kpc.
\end{enumerate}

We show $C$-index measured using methods (ii)-(iii) as a function of what is obtained using method (i) in Fig.~\ref{fig:cindexcompare}. Including or not a limit on the scaleheight (methods (ii) vs (i)) has very little impact on the resulting $C$-index. Using a spherical aperture of 100kpc to measure $C$-index (method (iii)) gives very similar results to method (i) for $C$-index values $<0$, while at higher $C$-index values the scatter increases in a way that method (iii) tends to give a higher $C$-index value. Because our interest is mostly on $C$-index values $<0$ and so we can more directly compare to SAMI, thus if we were to adopt method (iii) our results would not be affected. The comparison above shows that adopting a different method of measuring $C$-index has little impact on our result.

\section{Particle number selection} \label{Esec:sfrnum}

\begin{figure}
	\includegraphics[width=\columnwidth]{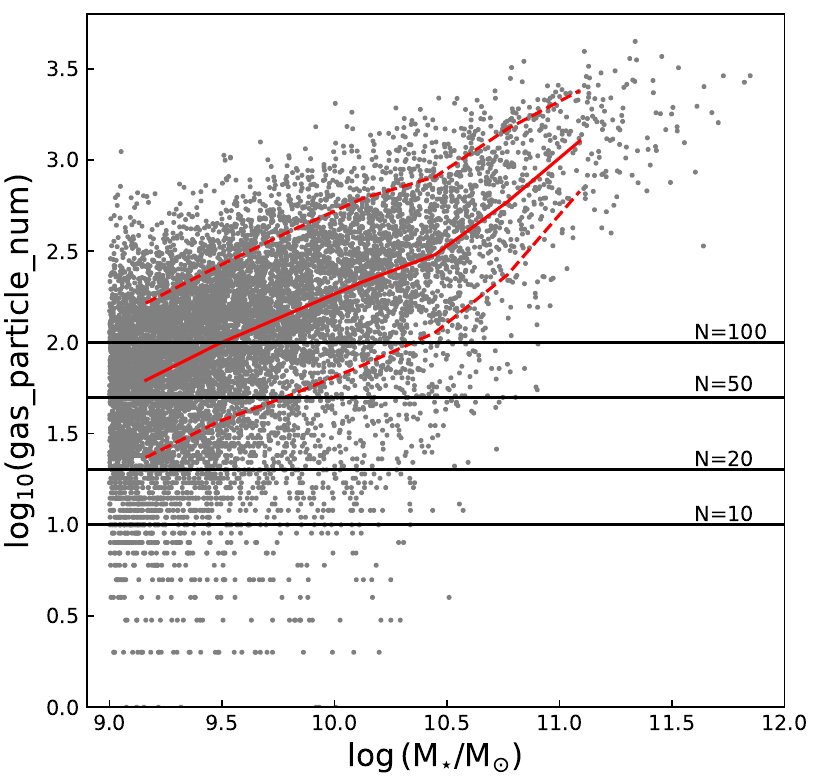}
    \caption{Number of gas particles with $\rm SFR>0$ versus $M_{\star}$ for all galaxies at $z=0$ with stellar masses $\ge 10^9\,\rm M_{\odot}$ that are within the cylinder aperture applied in method (i) (see text). The solid and dashed lines show the median and 1$\sigma$ percentile ranges. Horizontal lines highlight where numbers = 10, 20, 50, and 100 are for guidance. For the  galaxy sample used in this paper, we apply a cut on gas particles number > 20.}
    \label{fig:numsfr}
\end{figure}

\begin{figure}
	\includegraphics[width=\columnwidth]{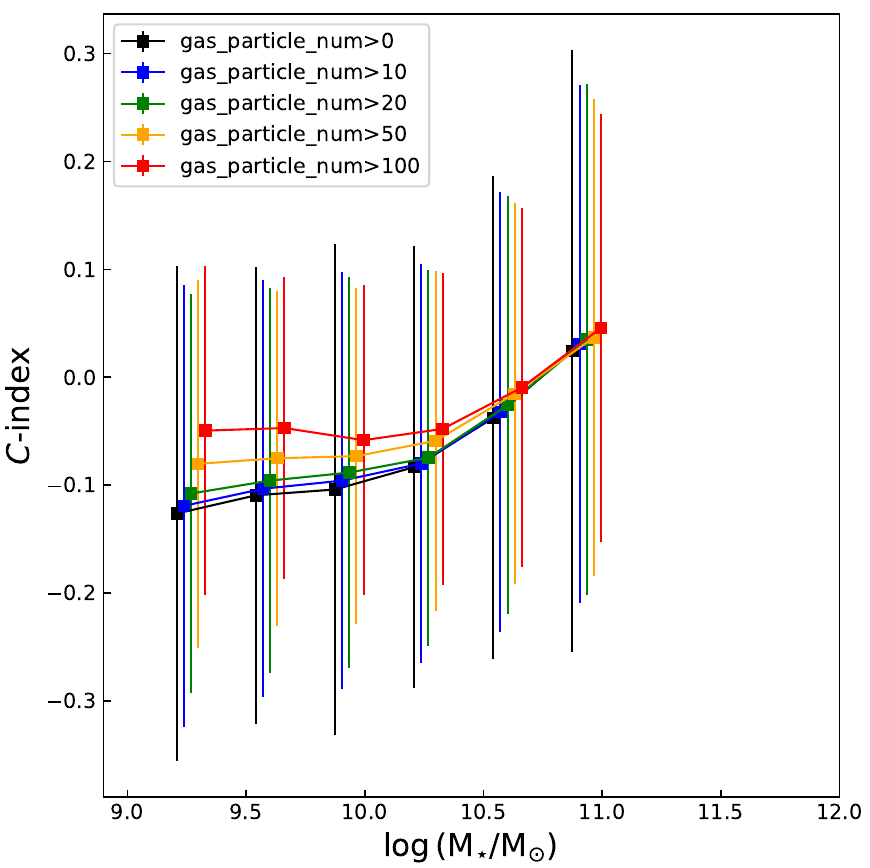}
    \caption{Median $C$-index in $M_{\star}$ bins for different cut on number of gas particles. The solid lines and errorbars lines show the median and 1$\sigma$ percentile ranges. Including low gas\_particle\_num galaxies will increase the range of scatter.  }
    \label{fig:C_vs_M_numsfr}
\end{figure}

\begin{figure}
	\includegraphics[width=\columnwidth]{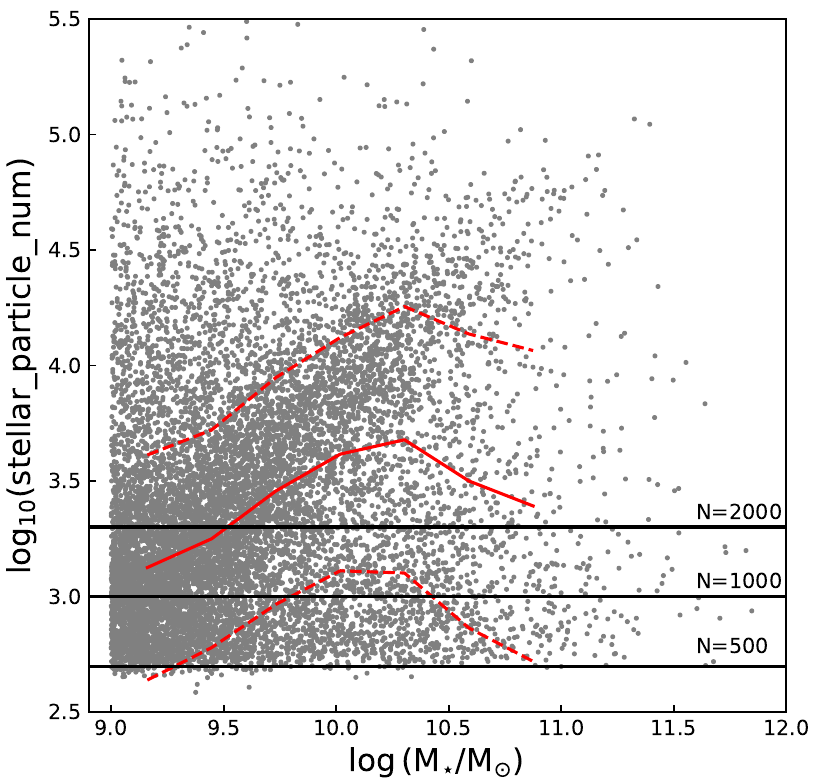}
    \caption{Number of stellar particles versus $M_{\star}$ for all galaxies at $z=0$ with stellar masses $\ge 10^9\,\rm M_{\odot}$ that are within the cylinder aperture applied in method (i) (see text). The solid and dashed lines show the median and 1$\sigma$ percentile ranges. Horizontal lines highlight where numbers = 500, 1000, and 2000 are for guidance. Because all galaxies have several hundred stellar particles, we do not apply a specific stellar particle number threshold to select the galaxy sample used to measure $C$-index.}
    \label{fig:numrband}
\end{figure}
To build a selection of galaxies to be used in our work, we first apply a stellar mass cut of $\ge 10^9\,\rm M_{\odot}$. This is motivated by the fact  that \citet{Furlong2015} showed that galaxies above that mass have SFRs that are converged in {\sc Eagle}. 

Secondly, and as mentioned in the previous section, we use gas and stellar particles to calculate $C$-index. We need to have a sufficient number of gas and stellar particles to perform the calculation. To understand how that impacts the selection of galaxies, we show the number of gas particles with $\rm SFR>0$ and the number of stellar particles within the cylinder described in method (i) in Appendix~\ref{Esec:cindex_compare}, versus $M_{\star}$ in Fig.~\ref{fig:numsfr} and Fig.~\ref{fig:numrband}, respectively. By applying a stellar mass cut, we are left with galaxies that have $\gtrsim 500$ of stellar particles within the defined cylinder, which is comfortably high enough to measure a half-light radius in the r-band. For gas particles, we see that there is a tail of galaxies with very few gas particles $<10$, but the $16^{\rm th}$ percentiles sit above that, at $\approx 30$ particles. {In Fig.~\ref{fig:C_vs_M_numsfr}, we show the median $C$-index versus $M_{\star}$ for galaxies using different cuts on the number of gas particles with SFR$>0$ inside the cylinder defined in (i) above. Applying different cuts on the number of gas particles does not change the main relation when the cut goes from $0$ to $20$, but the scatter is clearly larger when we allow too few gas particles to be used to measure $C$-index. The latter is due to the measurement becoming way too noisy in those cases. On the other end, if we are too conservative, and require $>50$ gas particles with SFR$>0$ in the cylinder to measure $C$-index, we start to bias the main relation due to removing the low SFR galaxies.} To avoid significantly biasing the selected galaxy population (which could happen if we chose a value for the number of gas particles that is too large), and to allow for enough gas particles to measure a half-SFR radius, we adopt a threshold of $20$. 


\bsp	
\label{lastpage}
\end{document}